\documentclass[preprint2]{aastex}

\def\pph{{p_{ph}^{}}} \def\rph{{\rho_{ph}^{}}}
\def\Tph{{T_{ph}^{}}} \def\Hph{{H_{ph}^{}}}
\def\Bph{{B_{ph}^{}}} \def\timph{{t_{ph}^{}}}

\def\Fz{{F_{z}^{}}}
\def\Flz{{F_{l_z}}}
\def\Fg{{F_{g}^{}}}
\def\Fp{{F_{p}^{}}}

\newcommand{\Eq}[1]{Eq.~(\ref{#1})}

\newcommand{\Fig}[1]{Fig.~\ref{#1}}

\newcommand{\Sec}[1]{Section~\ref{#1}}

\newcommand{\EQ}{\begin{equation}}
\newcommand{\EN}{\end{equation}}
\newcommand{\EQA}{\begin{eqnarray}}
\newcommand{\ENA}{\end{eqnarray}}
\newcommand{\UQ}{\begin{displaymath}}
\newcommand{\UN}{\end{displaymath}}
\newcommand{\UQA}{\begin{eqnarray*}}
\newcommand{\UNA}{\end{eqnarray*}}

\def\d2dt2{{{d^2} \over {dt^2}}}

\def\BB {{\bf  {B}}}

\def\Aa {{\bf  {A}}}

\def\rr {{\bf  {r}}}

\def\dbbz {{{\partial {{B^2}}} \over {\partial z}}}
\def\dPz {{{\partial {{P}}} \over {\partial z}}}

\def\dx2 {{{\partial ^2} \over {\partial x^2}}}
\def\P4  {{{\pi } \over {4}}}
\def\p4  {{{\pi } \over {4}}}

\def\ddt#1  {{{\partial #1} \over {\partial t}}}
\def\ddr#1  {{{\partial #1} \over {\partial \rr}}}
\def\ddx#1  {{{\partial #1} \over {\partial x}}}
\def\ddy#1  {{{\partial #1} \over {\partial y}}}
\def\ddz#1  {{{\partial #1} \over {\partial z}}}

\def\pmbf#1{\setbox0=\hbox{#1}\kern-.025em\copy0\kern-\wd0
            \kern.05em\copy0\kern-\wd0\kern-.025em\raise.0233em\box0 }

\def\Alfvenic-{Alfv{\'e}nic\ }
\def\Alfvenic{Alfv{\'e}nic }

\def\ltape{\hbox{\raise 1.5truept \rlap{$<$}\lower 3truept\hbox{$\sim$}\ }}
\def\gtape{\hbox{\raise 1.5truept \rlap{$>$}\lower 3truept\hbox{$\sim$}\ }}

\slugcomment{}

\shorttitle{Magnetic Flux Emergence}
\shortauthors{Galsgaard et al.}

\newcommand{\Tab}[1]{Table~(\ref{#1})}
\def\pph{{p_{ph}^{}}} \def\rph{{\rho_{ph}^{}}}
\def\Tph{{T_{ph}^{}}} \def\Hph{{H_{ph}^{}}}
\def\Bph{{B_{ph}^{}}} \def\timph{{t_{ph}^{}}}

\usepackage{color}

\begin{document}

\title{The effect of the relative orientation between the coronal field
and new emerging flux: I Global Properties}

\author{K. Galsgaard\altaffilmark{1}, V. Archontis\altaffilmark{4}, 
F. Moreno-Insertis\altaffilmark{2,3}  and A. W. Hood\altaffilmark{4}}
\altaffiltext{1}{Niels Bohr Institute, Julie Maries vej 30, 2100 Copenhagen
\O,  Denmark}
\altaffiltext{2}{Instituto de Astrofisica de Canarias (IAC),  Via
Lactea s/n, 38200 La Laguna  (Tenerife), Spain}
\altaffiltext{3}{Department of Astrophysics, Faculty of Physics, Universidad
de La Laguna,  38200 La Laguna (Tenerife), Spain}
\altaffiltext{4}{School of Mathematics and Statistics, University of St
Andrews, North Haugh, St Andrews, Fife KY16 9SS, UK}

\begin{abstract}
The emergence of magnetic flux from the convection zone into the corona is
an important process for the dynamical evolution of the coronal magnetic field.
In this paper we extend our previous numerical investigations, by looking at
the process of flux interaction as an initially twisted flux tube emerges
into a plane parallel, coronal magnetic field. Significant
differences are found in the dynamical appearance and evolution of the 
emergence process depending on the relative orientation between the rising 
flux system and any preexisting coronal field.
When the flux systems are nearly anti-parallel, the experiments show 
substantial reconnection and demonstrate clear signatures
of a high temperature plasma located in the high velocity outflow regions extending
from the reconnection region. However, the cases that have a more parallel orientation 
of the flux systems show very limited reconnection and none of the associated features.
Despite the very different amount of reconnection between the two flux systems, it is 
found that the emerging flux that is still connected to the original tube, reaches the same 
height as a function of time. As a compensation for the loss of tube flux, a clear 
difference is found in the extent of the emerging loop in the direction perpendicular
to the main axis of the initial flux tube. Increasing amounts of magnetic reconnection
decrease the volume, which confines the remaining tube flux.
\end{abstract}


\section{Introduction}
\label{introduction.sec}
Flux emergence is one of the manifestations of the continuously changing
solar magnetic field. In this process magnetic flux is transported up 
through the convection zone presumably by a combination of buoyancy and 
convection towards the photosphere. At the photosphere the physical structure 
of the sun changes, from a convectively unstable to a convectively stable 
atmosphere, making the continued rise of magnetic flux, due to buoyancy, 
more difficult. Despite this, numerous observations in all 
wavelength ranges show that the emergence process is a frequently 
occurring event in the solar atmosphere.

There are
numerous observations of the photosphere and lower transition region 
\citep{Lites_ea95,Lites_ea98,strouse_zwaan99,Kubo_ea03,Pariat_ea04,lites_05}.
These show a picture whereby emerging magnetic fields change the convective flows, 
allowing for initially horizontal magnetic fields to penetrate the photosphere. 
As the magnetic field expands into the transition region and lower corona, 
relatively cold plasma is lifted by the magnetic field and eventually starts 
draining back towards the photosphere, along the magnetic field lines.
In isolated emergence events two strong opposite polarities arise forming
two magnetically connected sunspots. In the region between the two flux 
concentrations new flux continues to emerge, with new positive - negative pairs 
arising at different positions. \cite{Lites_ea95} used observations
covering temperatures from photosphere to corona plasma to
argue that a full coherent flux tube rises from the convection zone.
\cite{strouse_zwaan99} suggested that the flux emerges in "parallel" sheets
between the two major flux concentrations, and that the random appearance of new small flux
concentrations between the two sunspots favor a picture whereby undulating
field lines emerge at different locations along the full length of the field line 
connecting the two sunspots. \cite{Pariat_ea04} 
uses field extrapolation of the photospheric field to establish a 
field line structure of the magnetic field and find that undulating field lines
exist in this model. Furthermore, the dips are associated with Ellerman bombs 
\citep{Georgoulis_ea02}, and they interpret these as being responsible for leaving
dense material in the part of the loop that dips below the photosphere, 
making the emergence for the remaining upper part of the loop much easier.

The manifestation of flux emergence in the solar corona often results in a strong 
interaction between the two initially disconnected flux systems. This interaction 
results in local brightenings observed in all wavelengths from white light through 
EUV and into X-ray.  Often these events are associated with high-speed flows of the
hot plasma emerging from the (reconnection) region where the two flux systems interact.
\cite{Longcope_ea05} investigated one such event. Using simple flux estimates for 
the flaring structures, they showed how the new flux regions changed its connectivity 
with time and became connected to the existing coronal magnetic field. This happened
in such a way that no dynamic activity took place for a initial period of time,
followed by a rapid change of connectivity, and finally a more quiet phase.

How does this interaction depend on the structure of the two
initially separate flux systems? No clear analysis has been made on 
existing observational data, and one can ask if this is actually possible, 
with the present inability of directly tracking magnetic field 
lines and their photospheric connections. Further to this, each observed
event is different from previous ones in a number of ways, so how 
can one quantify the most important reasons for these differences in evolution?

Such effects are much easier to investigate by undertaking a series of numerical 
experiments where one can control, in detail, the environment into which the
flux emergence occurs. 
In previous experiments 
\citep{fan01,magara+longcope03,archontisetal04,archontisetal05,manchesteretal04,magara06},
the rise of a twisted loop is initiated inside the convection zone and followed in a
self-consistent manner. In these investigations, the structure of the emerging flux 
resembles a situation where the outer layers of the flux tube expand into the corona. 
In a manner, this resembles the process of peeling off the outer layers of an onion and 
expanding these parts into the corona. Specifically \cite{magara06}, compared the 
development of the photospheric flux concentrations
with observations, finding the same characteristic evolution of the flux
concentrations and their associated inversion polarity line as seen in observations.
\cite{galsgaardetal05,archontisetal05} and \cite{Isobeetal05} presented 
the first 3D MHD experiments in which a twisted flux tube emerges from the
convection zone and interacts with a pre-existing coronal magnetic field. 

\cite{archontisetal05} analyzed in detail the interaction between an emerging magnetic flux 
system and a uniform horizontal coronal magnetic field, using for the rising tube the
same initial conditions as in previous experiments (e.g. \cite{fan01,archontisetal04}).
\cite{archontisetal05} show how the sub-photospheric flux tube emerges into the
corona and pushes the magnetic field upward and outward. Given the initial almost 
antiparallel, mutual orientation of the system at the time of first contact, a strong 
current sheet is formed at the interface. The interaction of the two flux systems then 
follows a complicated pattern that slowly changes in time. This interaction depends also 
on the relative orientation between the two systems, as has been pointed out 
by \cite{galsgaardetal05}.

In a different approach, \cite{fanandgibson04} used a 2D magnetic arcade field 
embedded in a constant temperature coronal region, into which they force a 
pre-defined curved loop structure using an imposed boundary electric field. 
This showed, for the setup with minimal reconnection between the two flux 
systems, that the twisted emerging loop entered into the "corona", eventually 
experienced a kink instability and strong currents were generated where reconnection
takes place due to non-ideal effects. The emergence of a twisted loop into the
corona is supported by \cite{Lites_ea95}. Therefore it is interesting to compare
the structure of current concentrations between \cite{fanandgibson04} and experiments
which included the full emergence process, knowing that differences in the coronal
structure and the emergence process may provide very different dynamical evolutions.

In this paper, we follow up on the work presented in \cite{galsgaardetal05} and 
\cite{archontisetal05}
and investigate, in more detail, the significant deviations in the emergence process as
the orientation between the two interacting magnetic flux systems is changed.
In \Sec{model.sec} we briefly describe the numerical model. Sections 3-6 contain 
the various results of the experiments and \Sec{discussion.sec} is a discussion
of the implications of these results. Finally, highlights of the results are
summarized in \Sec{summary.sec}.

\section{Model setup and numerical approach}
\label{model.sec}

The parameters of the magnetic flux tube, the background stratification and the 
initial conditions follow the work presented in \cite{galsgaardetal05} and 
\cite{archontisetal05}. As a reference for the following discussion we start by 
summarizing the initial state of the experiments and the numerical approach.

Our model consists of a highly stratified environment and a horizontally twisted 
magnetic tube. The background medium consists of an adiabatically stratified 
convection zone, an isothermal layer representing the 
photosphere, a region where the temperature steeply increases with 
height and represents the transition region and finally an isothermal 
layer with coronal temperatures. The tube center is located 
almost $2$ Mm below the base of the photosphere. The longitudinal component of 
the magnetic tube has a Gaussian profile with a central field strength of $3.8$ kG, 
while the twist is uniform around the axis of the tube.
This particular flux tube is stable towards the kink instability. 
This gives a plasma 
$\beta = 12.8$ at the axis of the tube. The rise of the tube is triggered by 
implementing a density deficit distribution that has a maximum at the middle 
of the axis of the tube and with a Gaussian distribution along the tube. 
Initially, a horizontal magnetic field is included in the atmosphere
above the lower transition region. The orientation of the ambient field relative to
the main axis of the tube is an important parameter in the experiments.

\begin{figure}[htb]
\centering \includegraphics[scale=.45]{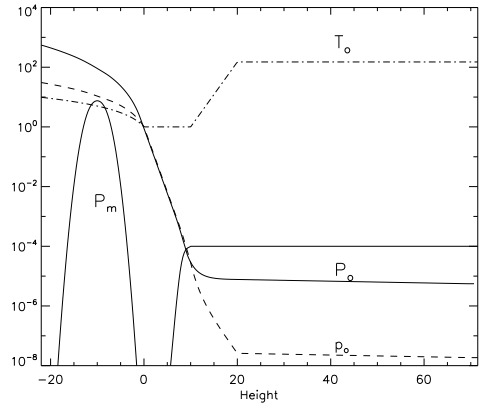}
\caption[]{
Distribution of gas pressure (thick solid), density (dashed), temperature
(dash-dotted) and magnetic pressure (thin solid) along the central, vertical 
($y=0$,$x=0$) line.
}
\label{init.fig}
\end{figure}

Figure \ref{init.fig} shows the gas pressure, temperature and density of the stratified 
environment as a function of height. All the profiles are normalized according to the 
photospheric values:  $\pph = 1.4\,10^5$ erg cm$^{-3}$;
$\rph = 3\,10^{-7}$ g cm$^{-3}$; $\Tph = 5.6\,10^3$ K and 
$\Hph = 170$ km. Other units used in the simulations are:
time, $\timph = 25$ sec; velocity, $V \equiv (\pph/\rph)^{1/2}
= 6.8$ km sec$^{-1}$ and magnetic field, $\Bph =1.3\,10^3$ Gauss.

The distribution of the magnetic pressure in \Fig{init.fig} shows the magnetic flux 
tube and the ambient field. The ambient field is given by 
\EQ
{\bf B}_{cor} = B_c(z)\;[\cos(\phi),\sin(\phi),0],
\label{eq:coronaprofile}
\EN
where $B_{c}(z)$ is described by an hyperbolic tangent profile. The intensity 
of the coronal field is chosen such that the local plasma $\beta$ is close 
to $0.06$.

\begin{figure*}[!htb]
\centering 
\includegraphics[scale=.20]{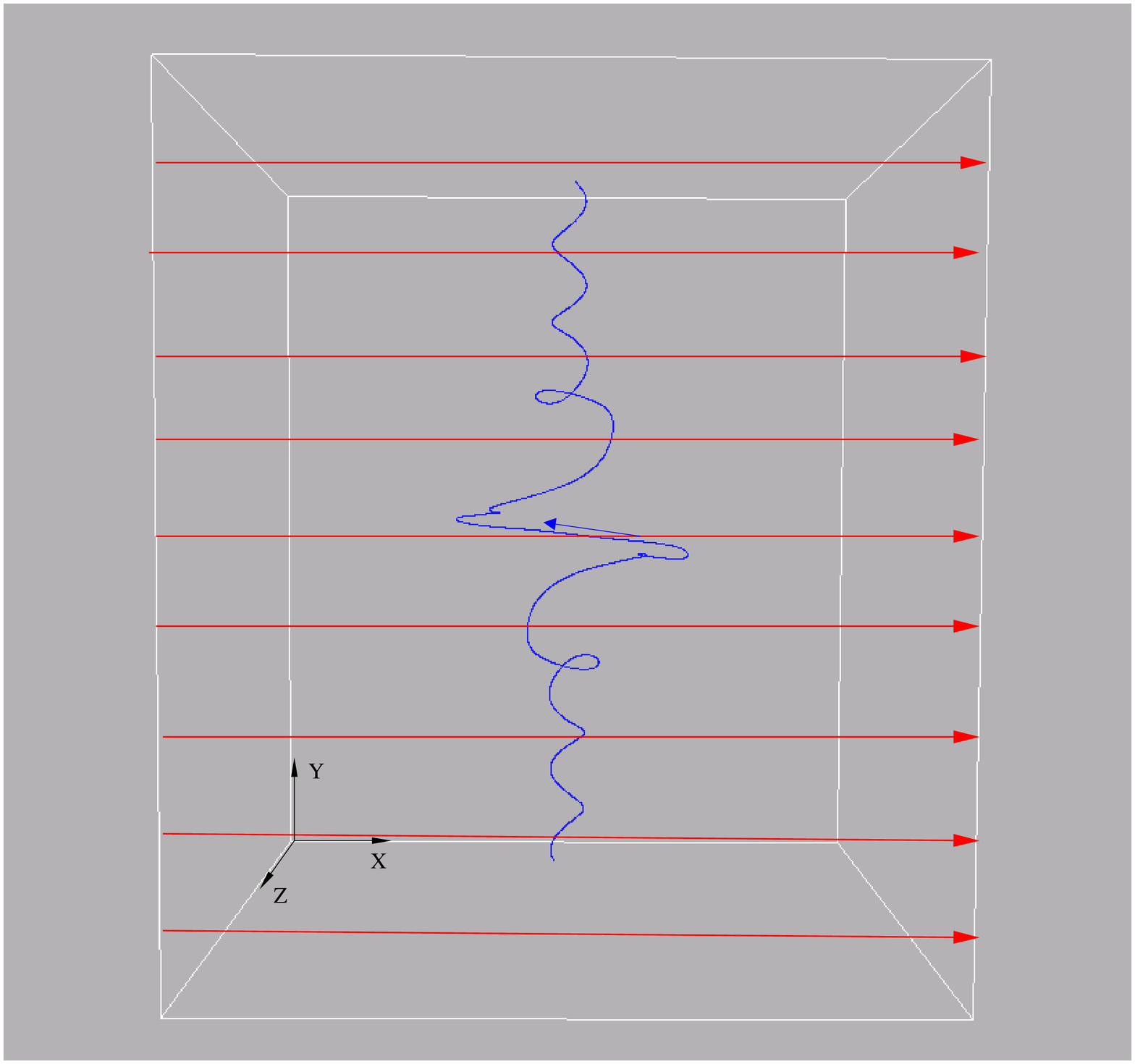}
\includegraphics[scale=.20]{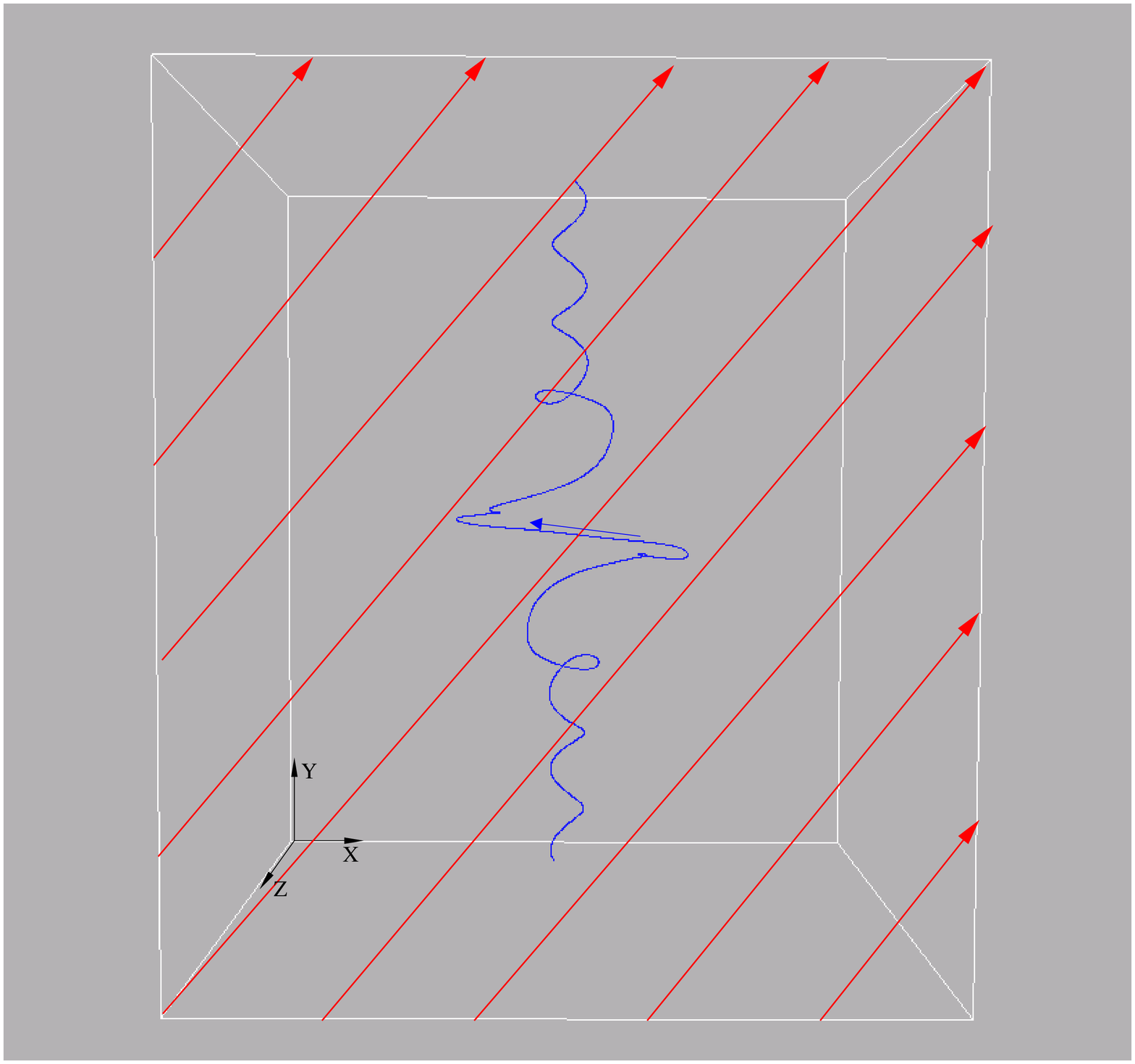}
\includegraphics[scale=.20]{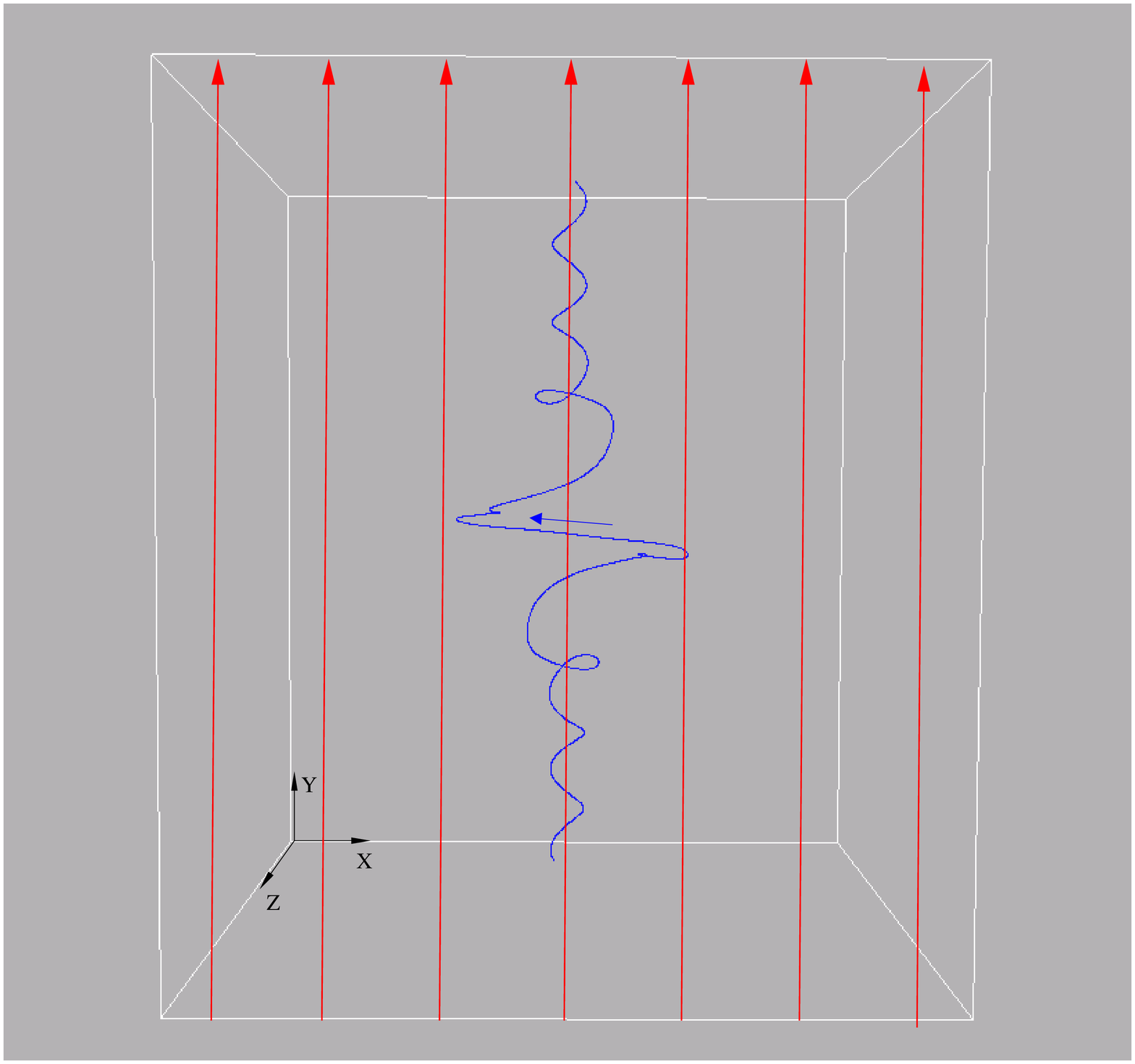}
\caption[]{Top view of an emerging twisted fieldline and 
the ambient coronal field (long horizontal arrows). The short arrow at the 
center of the box shows the direction of the emerging fieldline. The 
three panels represent different relative orientations of the two 
flux systems: experiment A: $\phi=0$, experiment B: $\phi=45$ 
and experiment C: $\phi=90$.
}
\label{orientation.fig}
\end{figure*}

The direction of the initial ambient field is given by the polar angle $\phi$, 
which is measured in a horizontal $xy$-plane from the positive $x$-axis. 
At the same time the magnetic fieldlines at the top of the rising flux system 
are oriented in an approximately antiparallel direction to the ambient fieldlines 
when $\phi=0$ and they are almost parallel when $\phi=180$ (see 
\Fig{orientation.fig}). Thus, the relative horizontal angle between the two flux 
systems upon contact is $\phi_0=180-\phi$ deg. The polar angle, $\phi$, of the coronal
magnetic field is different in the five experiments we have performed; it changes
from $\phi=0$ to $\phi=180$ (see \Tab{names.tab}). 

\begin{table*}[htb]
\begin{center}
\begin{tabular}{| c | c | c | c |} \hline \hline
Experiment & $\phi_0$  & Line style           & Orientation   \\ \hline
A          & 180.      & Triple Dotted dashed & Antiparallel  \\
B          & 135.      & Dot-Dashed           & Slanted       \\
C          & 90.       & Full                 & Perpendicular \\
D          & 45.       & Long-Dashed          & Slanted       \\
E          & 0.        & Dotted               & Parallel      \\ \hline
\hline
\end{tabular}
\caption[]{List of the numerical experiments performed in the simulations.
The first column shows the reference names of the experiments. The second column shows 
the relative horizontal angle of the two magnetic systems. 
The third column indicates the linestyle used in the different plots in this paper and the 
fourth column the orientation of the coronal field relative to the upcoming field.}
\label{names.tab}
\end{center}
\end{table*}

The evolution of the system is governed by the three-dimensional, time-dependent 
and resistive MHD equations. These are solved using a numerical approach based on high 
order finite differencing on staggered grids. By using 6 neighboring data points,
a 6th order accurate spatial derivatives and corresponding 5th order accurate interpolations
routines are used. The solution is advanced in time using a 3rd order predictor-correction
algorithm. Due to the high spatial order, special treatment of viscosity and resistivity are
required to prevent numerical ringing in the vicinity of steep gradients in the physical
quantities. This is handled by a combined approach, that is designed to remove numerical 
problems that can occur in specific problems.
These approaches are localized in space,
implying that dissipation only take place over length scales of a few gridpoints. Using such 
an approach, makes it impossible to assign a single characteristic Reynolds number for the 
experiment \citep{Nordlund+Galsgaard97}. 

The numerical resolution of the experiments is $(148,160,218)$ in the $(x,y,z)$ 
directions, with $z$ being the height. The size of the numerical domain is $(-60,60)$, 
$(-70,70)$ and $(-22,70)$, which is equivalent to a box of sides $20.4$ Mm x $23.8$ Mm 
x $15.6$ Mm. The resolution in the $x$ and $y$ directions is $137.8$ km/cell and 
$148.8$ km/cell correspondingly. The grid in the vertical direction is stretched in a way
that the highest resolution covers the region from the top of convection zone to the bottom
of the corona. Here the grid resolution is $47.7$ km/cell. The resolution has lower 
values close to the top and bottom boundary of the numerical domain.

\section{Current sheet orientation}
\label{current_sheet.sec}
The location, orientation and strength of current sheets in 3D is vital
for providing an environment for fast magnetic energy release. This section is concerned
with investigating the buildup of current sheets in the various experiments, with the aim to 
study the relation between some basic model parameters and locations of reconnection.

\begin{figure}[!htb]
\centering \includegraphics[scale=0.40]{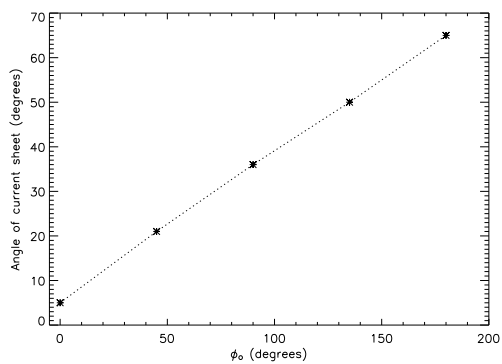}
\caption[]{The orientation of the current sheet with respect to the vertical $yz$ midplane 
as a function of the angle between the two magnetic flux systems.
 }
\label{fig:current_angle}
\end{figure}

As the emerging flux pushes its way into the coronal magnetic field, stress builds 
up at the interface between the two flux systems. 
When the two flux systems are antiparallel ($\phi_0=180$), a current concentration 
is formed all over the emerging plasma hill.
As the stress continues to build up the current is concentrated 
into a narrow curved sheet that reaches from the summit point of the plasma hill down 
its sides towards the photosphere almost along the direction of the underlying emerging
flux tube in the $y$-direction. As the orientation of the coronal magnetic field 
changes in the different experiments, the orientation and strength of the 
current sheet changes too. The current sheet is found to rotate around its vertical 
central axis as a monotonic function of the angle $\phi_0$ between 
the coronal magnetic field and the emerging magnetic field (see also \Fig{fig:current_angle}), 
where $\cos \phi_0 = ({\bf B}_{cor}\cdot {\bf B}_{t})/|{\bf B}_{cor}||{\bf B}_{t}|$,
${\bf B}_{t}$ represents the tube field at the summit point of the emergence region.

The orientation of the current sheet can be found using the 
following analysis. Assume we are only interested in the orientation at the summit
point of the emergence, then the $z$ component of the field can be ignored. In general,
the field vectors can be expressed in terms of an orthogonal coordinate system, 
with one unit vector, ${\bf e}_{1}$, defined by the direction given by the sum of 
the magnetic vectors, ${\bf B}_{cor} + {\bf B}_{t}$ and the other unit vector, ${\bf e}_{2}$,
orthogonal to this defining a right hand system. Thus,
\EQA
    {\bf B}_{cor} & = & B_{cor\parallel}{\bf e}_{1} + B_{cor\perp}{\bf e}_{2}, \\
    \qquad {\bf B}_{t} & = & B_{t\parallel}{\bf e}_{1} + B_{t\perp}{\bf e}_{2}
\ENA
The components of ${\bf B}_{cor}$ and ${\bf B}_{t}$ along ${\bf e}_{1}$, namely 
$B_{cor\parallel}$ and $B_{t\parallel}$, represent the components of the magnetic field 
that cannot be annihilated and at the same time determine the direction of the main 
axis of the current sheet. The components along ${\bf e}_{2}$, $B_{cor\perp}$ and 
$B_{t\perp}$, provide two oppositely directed components that can annihilate in a
reconnection process. The integrated current across the sheet is simply
given by $|B_{cor\perp}|+|B_{t\perp}|$. Visualization of the current sheet and illustration 
of its orientation in the computational volume is shown in Section \ref{reconnect.sec}.

\subsection{Magnetic Pressure Balance Across the Current Sheet}
\label{pressure.sec}
In 2D the total pressure balance across a current sheet is simple, and it maintains a change
between magnetic and gas pressure. In 3D such a balance may change with time as the
relative orientation of the field that is advected into the sheet changes. In this paper, the
evolution of the total pressure balance 
is used as an indirect indicator of magnetic reconnection.
\begin{figure*}[!htb]
\hbox to \hsize{\hfill
\includegraphics[scale=0.83]{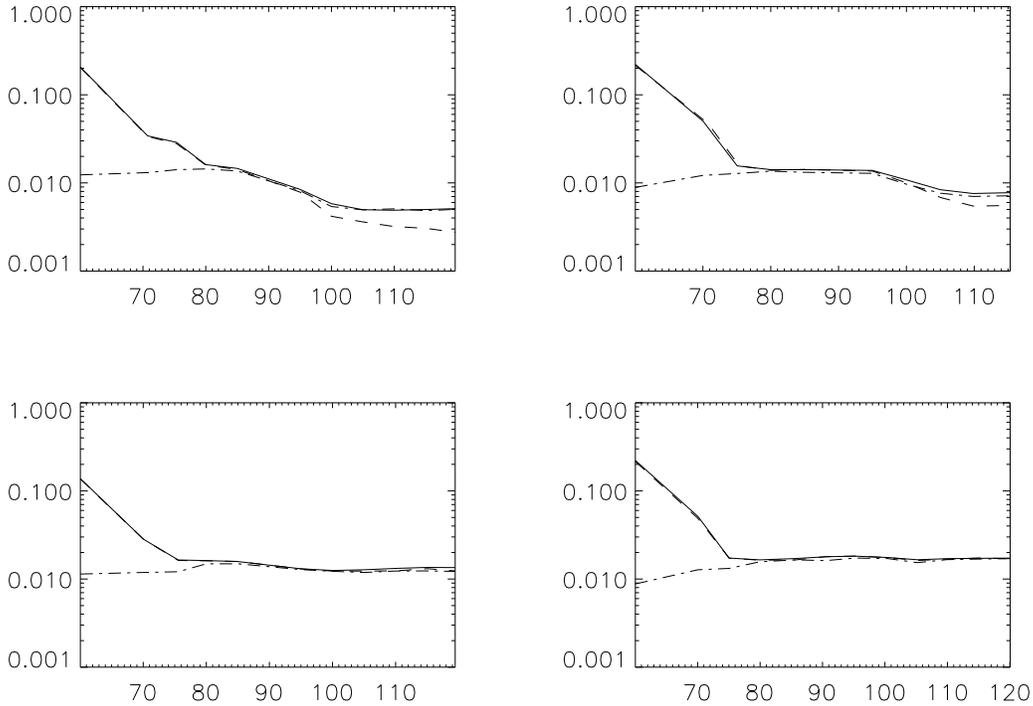}\hfill}
\caption[]{
Temporal evolution of the magnetic pressure just above and below the current sheet 
are shown for experiments A, B, C and D from top-left to bottom-right panel.
Full line represents the pressure below the sheet, the dot dashed line indicates
the pressure above the sheet. Finally the dashed line indicate the pressure from the
magnetic component perpendicular to the tube direction just below the sheet.}
\label{balance.fig}
\end{figure*}

As the tube rises the current sheet moves upward covering the upper part of the
buoyant flux system. \cite{archontisetal05} showed that the magnetic pressure distribution 
across the current sheet changes with time in experiment A. This change in the temporal 
evolution of the magnetic pressure was found to play an important role in the reconnection 
process between the two flux systems.

Figure \ref{balance.fig} shows the pressure of the horizontal component of the magnetic field 
($B_{hor}$) below (solid line, $|{\bf B}_{t}|$) and above the current sheet (dot-dashed line, 
$|{\bf B}_{cor}|$) as a function of time. The dashed line represents the pressure from the 
$B_x$ component of the magnetic field below the current sheet.

At the beginning of the emergence process the magnetic pressure inside the rising plasma is 
much larger than that of the ambient coronal field. At $t \approx 60$ the difference is more 
than one order of magnitude. This pressure excess pushes the emerging tube upwards into the 
atmosphere -  for details see the discussion of forces given in \Sec{forces.sec}.
After $t \approx 80$ 
a transverse balance of magnetic pressure is achieved, independently of the different orientation 
of the ambient field, and this balance remains until the end of the simulation.

Before $t \approx 95$ the $B_x$ and the $B_{hor}$ of the rising magnetic field in the tube,
are approximately equal. For experiments A and B the two components separate around this time, 
indicating that the orientation of the emerging magnetic field just below the interface changes 
with time. The reason being that, as time proceeds, the uppermost field lines of the rising 
plasma reconnect with the ambient field and this allows for different internal flux layers to come 
into contact with the overlying flux system; the magnetic field vector in these internal 
layers points increasingly away from the transverse direction and this explains the decrease 
apparent in the dashed curve in experiments A and B. On the other hand, for experiment C and D
the reconnection affects a much shallower region of the rising tube, and, as a result, the 
x-component of the field as the interface does not decrease.


Finally, it is found that pressure at the end of the experiments, $t = 120$, has an almost 
linear dependence on the value of $\phi$, providing the highest magnetic pressure for the cases
that do not show effective reconnection in \Fig{balance.fig}. The effects and importance
of reconnection is further discussed below.

\section{Magnetic Connectivity}\label{connect.sec}
In the previous section we showed that 
the emerging flux tube reconnects with the coronal magnetic field, at least for the
cases A - C. As a result 
of this the magnetic pressure below the current sheet changes with the relative 
horizontal angle $\phi_0$. Thus, it seems plausible that the emergence process 
will be strongly influenced by the orientation of the coronal field.
For example, when the two fields are approximately parallel one might 
expect the reconnection process to be slowed substantially and the emergence 
process possibly hindered. This section investigates the efficiency of reconnection,
from a global point of view, by 
\begin{itemize}
\item measuring the height of the apex and axis of the emerging tube in time (\Sec{height.sec}),
\item measuring the amount of horizontal and normal flux that emerges into the corona 
(\Sec{flux_emerg.sec}),
\item studying the changes in field line connectivity (\Sec{fieldline_connectivity.sec}), 
and finally by
\item measuring the fraction of the tube flux that reconnects in time 
(\Sec{flux_connectivity.sec}).
\end{itemize}

\subsection{Height-time relation: apex and axis}
\label{height.sec}

\begin{figure}[!htb]
\centering \includegraphics[scale=.48]{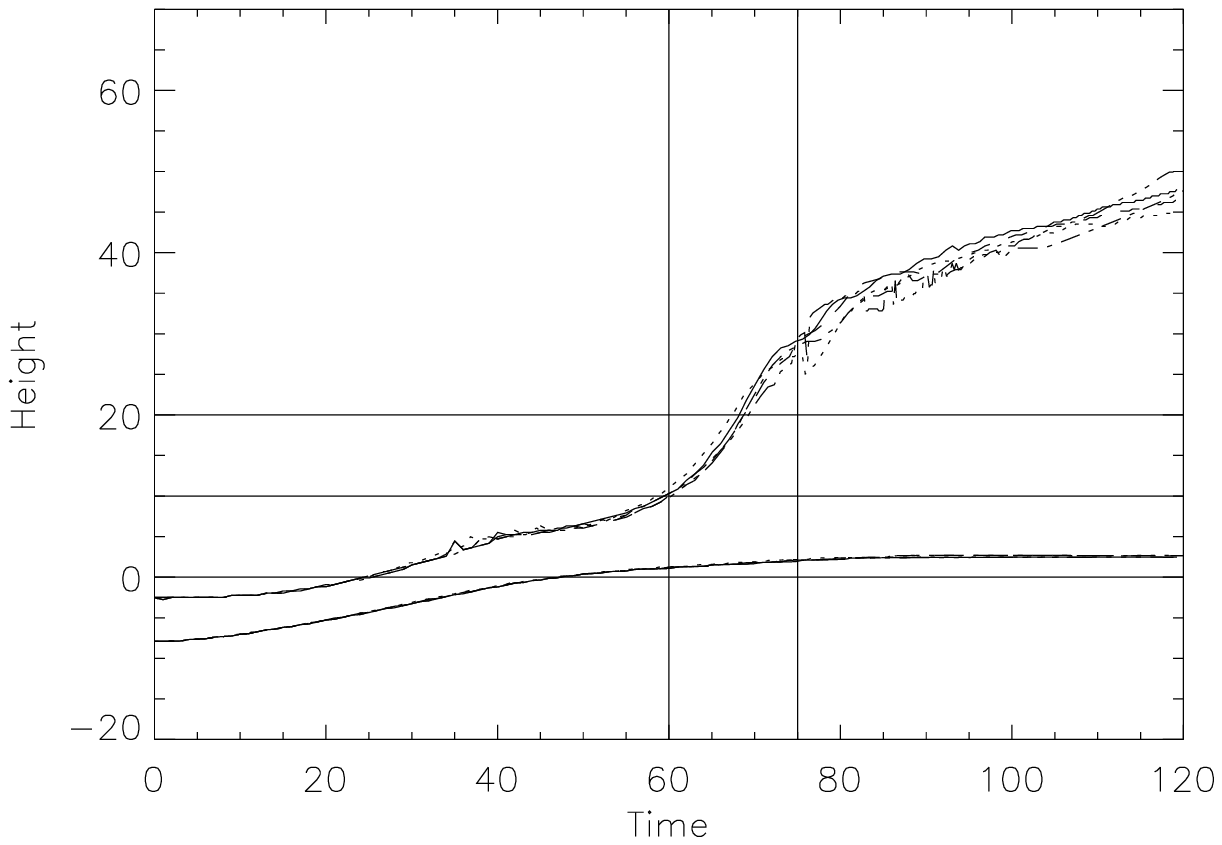}
\centering \includegraphics[scale=.48]{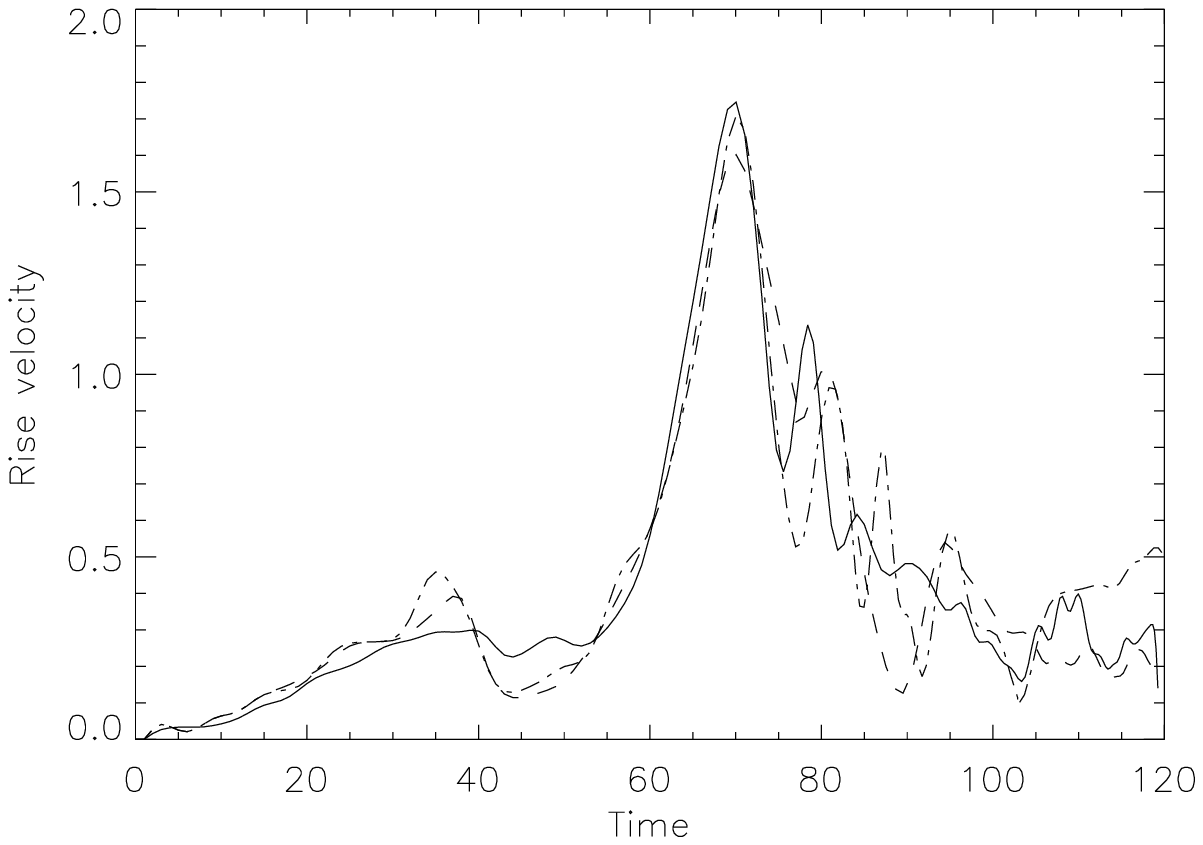}
\caption[]{\label{loop_height.fig}
Left: The height of the apex (upper curves) and the center (lower curves) of the tube 
as a function of time. Right: The associated velocity of the apex of the rising system.
For line styles see Table \ref{names.tab}.
}
\end{figure}

To measure the dependence of the rising motion of the tube on the orientation of the 
ambient field, we find the height of the apex and the axis of the rising tube as 
a function of time for the experiments listed in Table \ref{names.tab}.
More precisely, a large number of field lines are traced from starting points 
along the central vertical line. Then we find those fieldlines that 
stay in the tube and those that belong to the ambient magnetic field. 
The summit point is then the first point along the central line at which the 
connectivity changes.

\Fig{loop_height.fig} (left panel) shows 
the height-time relation of the apex and the center of the emerging tube.
The emergence starts with a slow rise phase while the flux tube is below the photosphere
(\cite{murray_ea06} investigate how this initial emergence phases depend on the 
tube parameters). This is followed by a rapid rise phase between $t=55$ and $t=80$ during 
which the apex of the tube rises through the transition region and into the corona. 
After $t=80$ the rise rate slows down and settles down to a lower rate that
fluctuates with time and between the various experiments, but on average follows
the same trend. The change in the height-time relation around $t=75$ corresponds to 
the time at which magnetic pressure balance across the current sheet is achieved, as 
seen in \Sec{pressure.sec}. At this time the width of the current sheet shrinks to the 
numerical resolution limit.

The lower curves in the top panel of Figure \ref{loop_height.fig} show that the axis 
of the tube reaches the photosphere at $t \approx 45$ and remains close to the
lower region of the photospheric layer until the end of the simulation.
A similar result is found in previous experiments of flux emergence
\citep{fan01,magara+longcope03,archontisetal04,manchesteretal04}.

Finally, the right panel of \Fig{loop_height.fig} shows the rise velocity of the apex
of the tube. 
It is seen that all experiments follow the same evolution until $t\approx 75$, after which a 
general decrease with superimposed oscillations are seen in the rise velocity. 

\subsection{Emerging flux}
\label{flux_emerg.sec}

\begin{figure}[!htb]
\centering \includegraphics[scale=0.48]{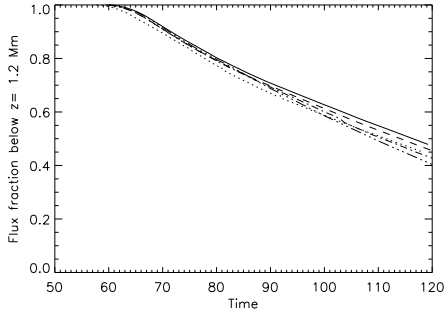}
\caption[]{
The graphs show the fraction of flux below $z=1.2$ Mm as a function of time for the five 
experiments mentioned in \Tab{names.tab}. Almost 65\% of the flux has emerged into the 
corona at t=120 for experiment A.
}
\label{emerge.fig}
\end{figure}

Another global measurement of the amount of the emerging flux in time can be 
obtained by calculating the amount of new horizontal flux passing through the 
vertical midplane ($x$-$z$ plane located at $y=0$) and above a height of $z=1.2 Mm$.
This particular reference height is chosen to be above the initial flux tube
and below the initial coronal magnetic field. In this way, the initial
coronal flux, independent of its orientation, does not contribute to the measurement. 

Alternatively, we can measure the amount of horizontal flux that remains below the 
height of $z=1.2 Mm$ in time. This amount of flux is defined by:
\EQ
  \Phi(t) = \int_{-L_{x}}^{L_{x}}\int_{-L_{z}}^{1.2}B_{y}(x,0,z,t) dz dx.
\EN
\Fig{emerge.fig} shows 
$\Phi(t)$, normalized by its value at $t=0$, as a function of time for the five experiments.
The profiles for the five experiments are almost identical until $t=90$.
This result shows that the amount of flux that emerges into the upper atmosphere is almost 
independent of the relative horizontal angle, $\phi_0$. After
$t=90$, we find that the larger the initial angle, $\phi_0$, is between
the emerging flux and the coronal magnetic field, the
larger is the fraction of the magnetic flux that emerges into the coronal regime at a
given time. However, the difference in emerged flux between experiment A and experiment 
E is found to be small, close to 7\%, at the end of the experiment, at which time
about 60\% of the initial flux has emerged into the outer atmosphere.

Observationally \citet{Kubo_ea03,Spadaro_ea_04} and \citet{Zuccarello_ea_05} have measured 
the development of the normal flux represented by the emerging region. They find a
time dependent growth, that over the first few days is close to linear. This is followed
by a saturation of flux and eventually a decrease. The grow rates of the magnetic flux in the
three cases are naturally different. One striking difference with the numerical 
experiments is the time scale involved. In the observations the timescale of the evolution of the system is measured in 
days, while the experiment here only covers about 20 minutes. Therefore it will not be 
possible to make a real comparison, but it is interesting to compare the structure of 
the comparable numerical measurement of the emerged magnetic flux.

\begin{figure}
\centering
\includegraphics[scale=0.48]{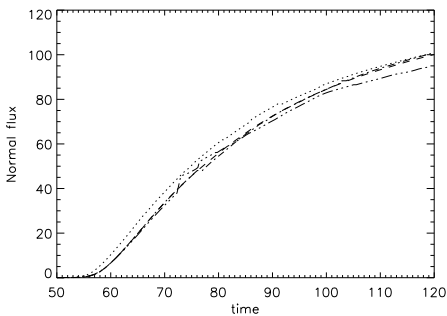}
\caption[]{\label{emerge_norma.fig} Time evolution of the normal flux
of the emergence through the $z=1.2$ Mm plane for the five 
experiments. For line styles see \Tab{names.tab}.
}
\end{figure}

If one assumes the emerging region is at the disk center,
then the observed quantity is equivalent to integrating the positive(/negative) flux
represented by the $B_z$ component in our experiments: 
\EQ
  \Phi(t) = \int_{-L_{x}}^{L_{x}}\int_{-L_{y}}^{L{y}}B_{z}(x,y,z,t) dy dx.
\EN
\Fig{emerge_norma.fig} shows the time evolution of the positive vertical flux through 
the transition region ($z=1.2 Mm$). The plot shows two different phases in the evolution: 
first a near linear increase of the vertical flux until about $t=75$, followed by another 
phase with a continuously decreasing rate. The difference in the time evolution of the 
vertical flux between the different experiments is very small.

Despite significant differences in the timescales between observations and these experiments,
the basic structure of the emerging flux appears very similar.

\subsection{Field line connectivity}
\label{fieldline_connectivity.sec}
The magnetic energy released in reconnection events is bound to spread along the reconnected field lines
due to anisotropic heat conduction. The bright structures found in EUV
and X-ray observations in association with flux emergence, and other energy release events,
are therefore providing vital information about the field line connectivity in the corona.
Comparing the coronal structures with models provides us with a possibility of understanding 
the field line structure of dynamical events.

There are several ways to show, in a qualitative manner, how the fieldline connectivity 
changes in time for the various experiments. We choose to trace field lines starting 
from one end of the submerged tube and see if they either connect
to the other end of the tube or to the corona. A disk is selected at the tube end, 
centered on the initial tube axis, and the destination of a large number of field lines 
is determined. Field lines going from the one end of the tube to the other are colored grey 
and field lines connecting to the corona are colored black. This method indicates the global 
connectivity of the fieldlines between the two flux systems. Details of the method can be 
found in \cite{Parnell+Galsgaard04}.

\begin{figure*}[!htb]
\hbox to \hsize{\hfill
\includegraphics[scale=.35]{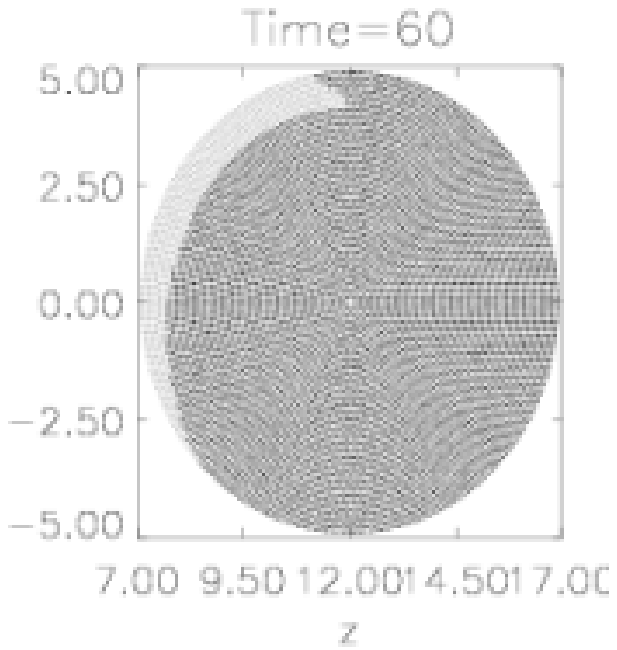}
\includegraphics[scale=.35]{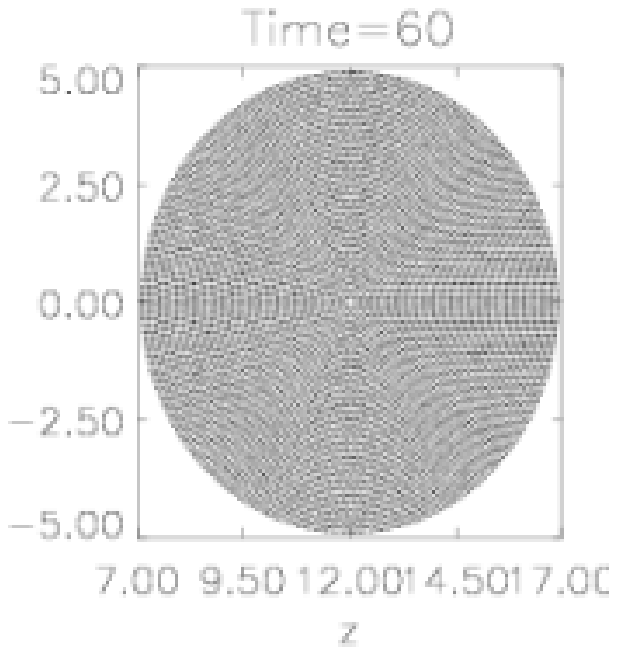}
\includegraphics[scale=.35]{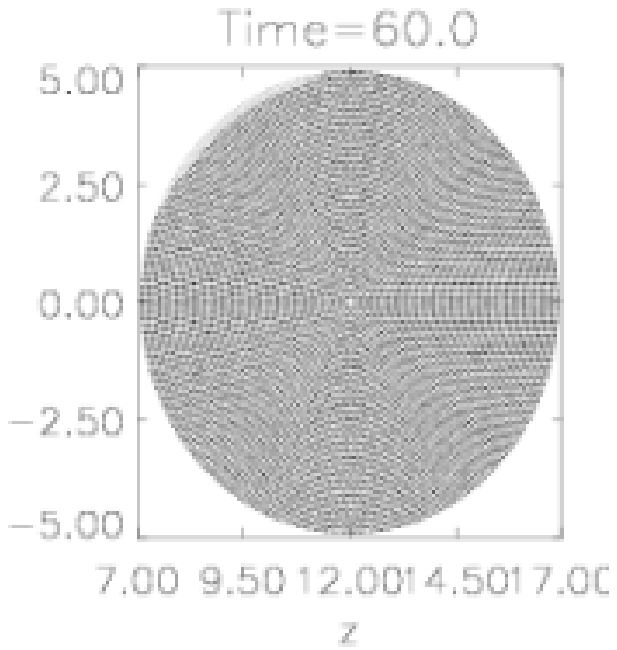}
\includegraphics[scale=.35]{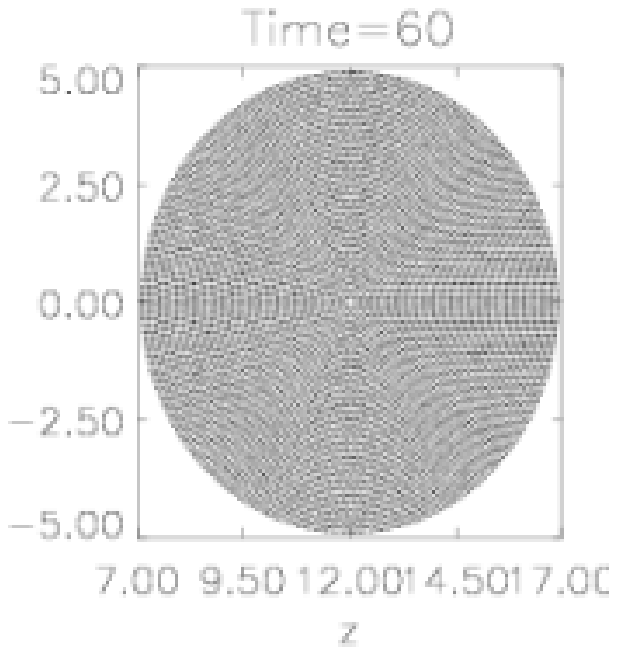}
\includegraphics[scale=.35]{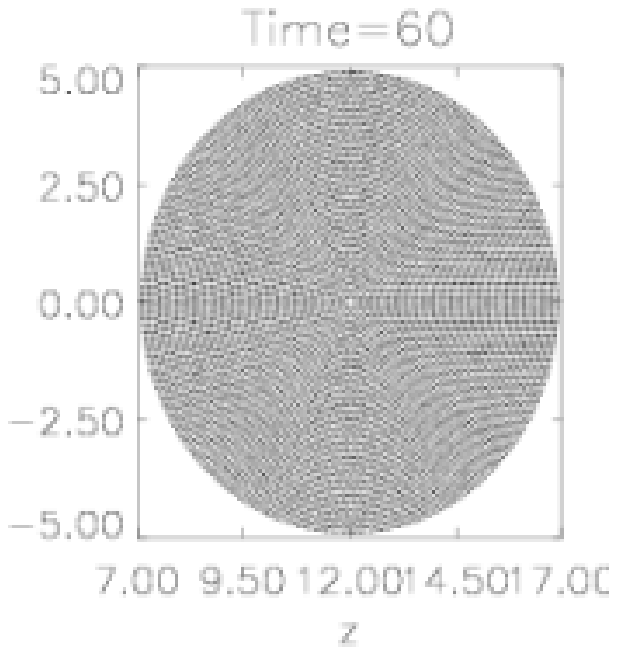}\hfill}
\hbox to \hsize{\hfill
\includegraphics[scale=.35]{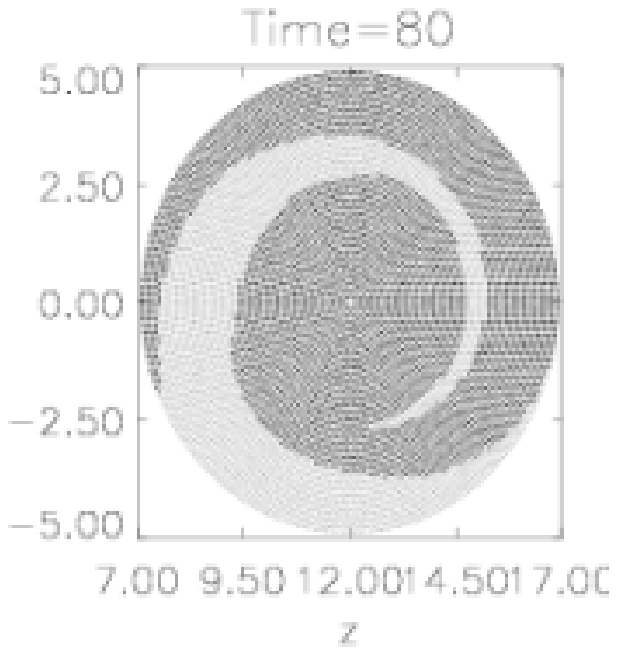}
\includegraphics[scale=.35]{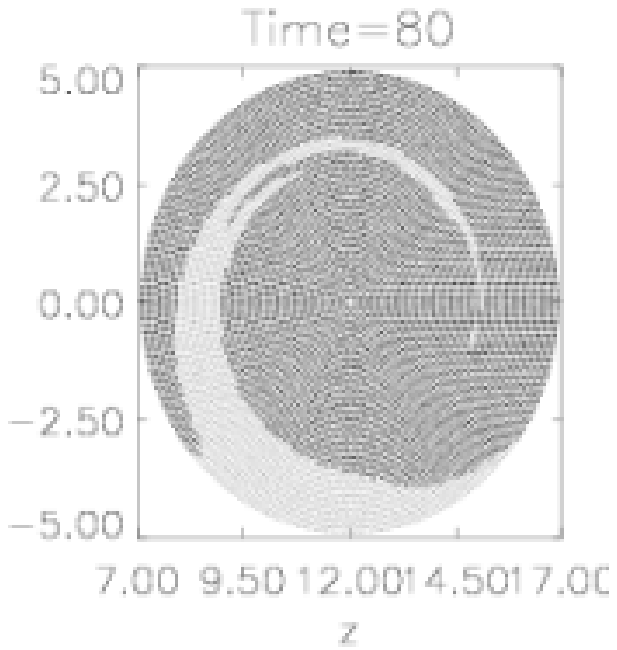}
\includegraphics[scale=.35]{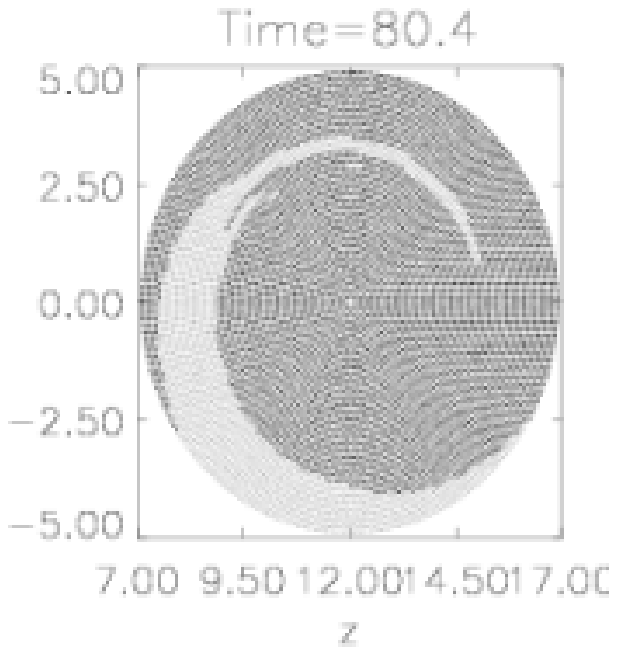}
\includegraphics[scale=.35]{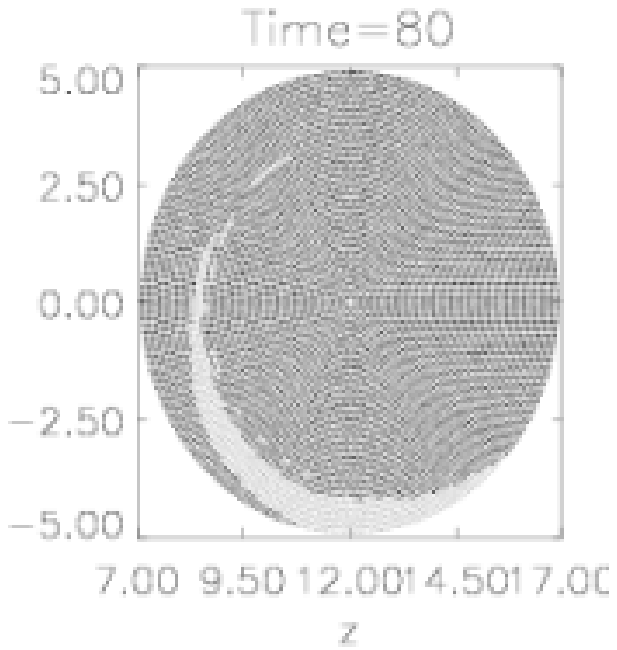}
\includegraphics[scale=.35]{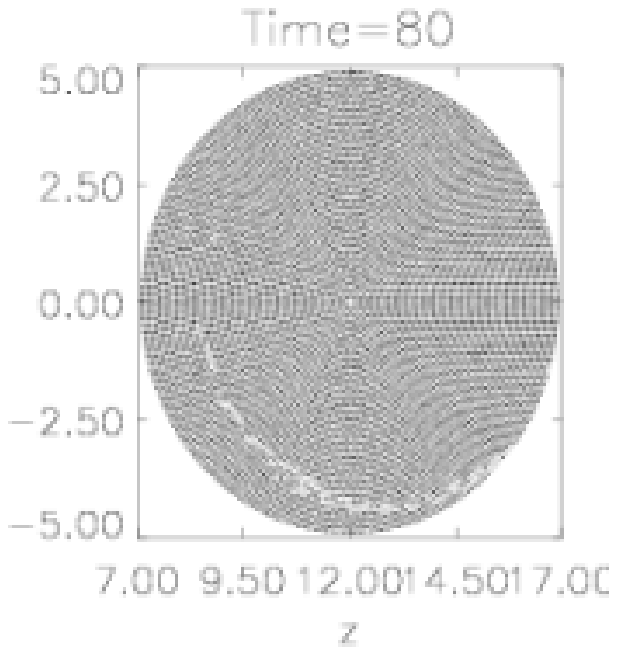}\hfill}
\hbox to \hsize{\hfill
\includegraphics[scale=.35]{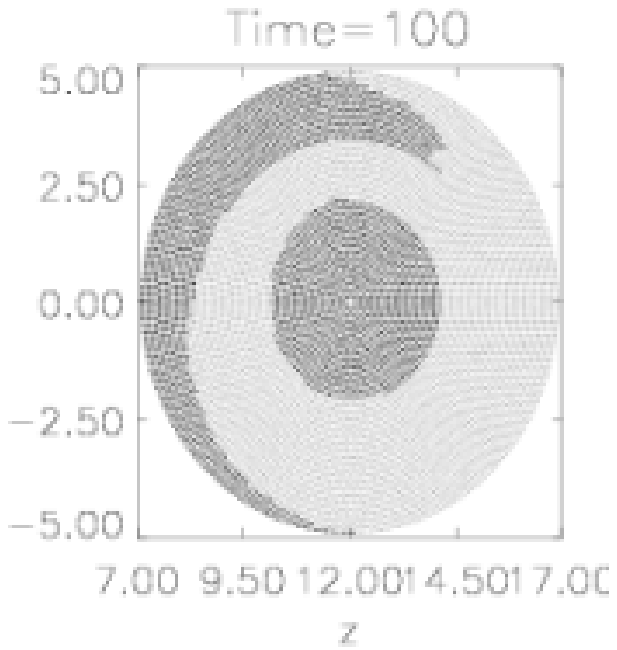}
\includegraphics[scale=.35]{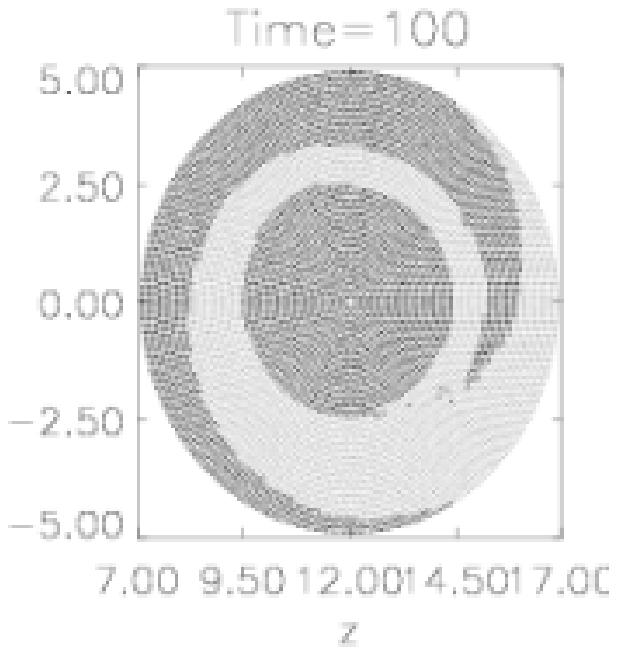}
\includegraphics[scale=.35]{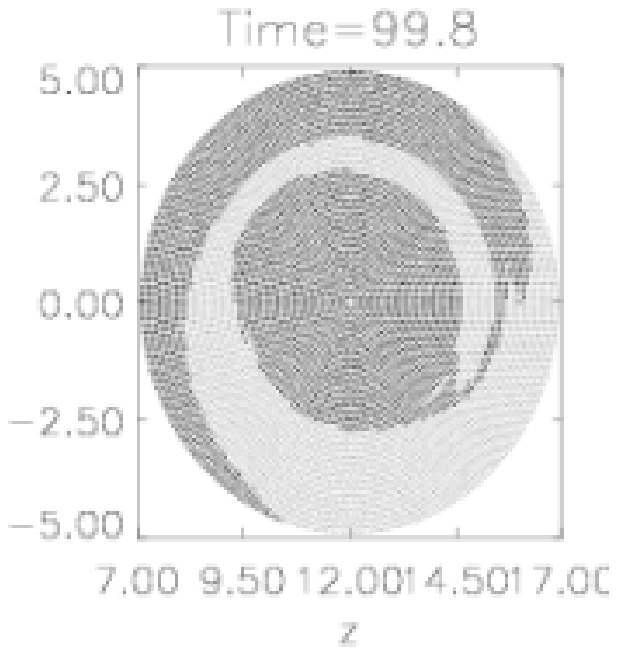}
\includegraphics[scale=.35]{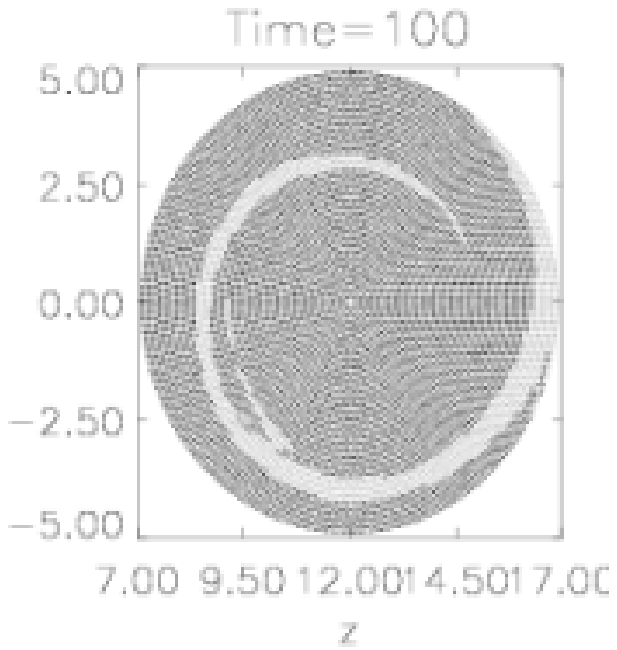}
\includegraphics[scale=.35]{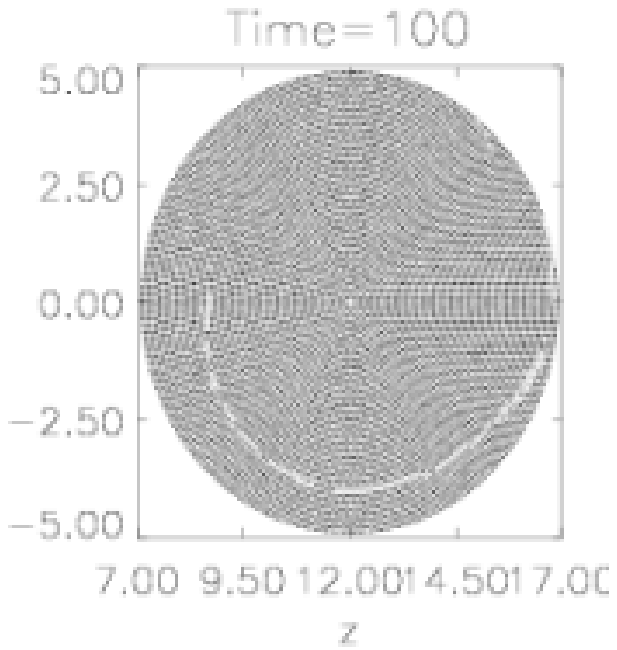}\hfill}
\hbox to \hsize{\hfill
\includegraphics[scale=.35]{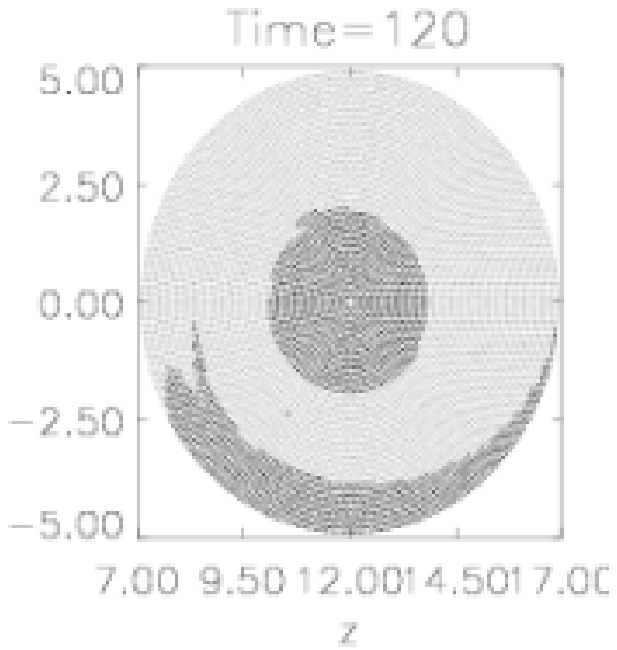}
\includegraphics[scale=.35]{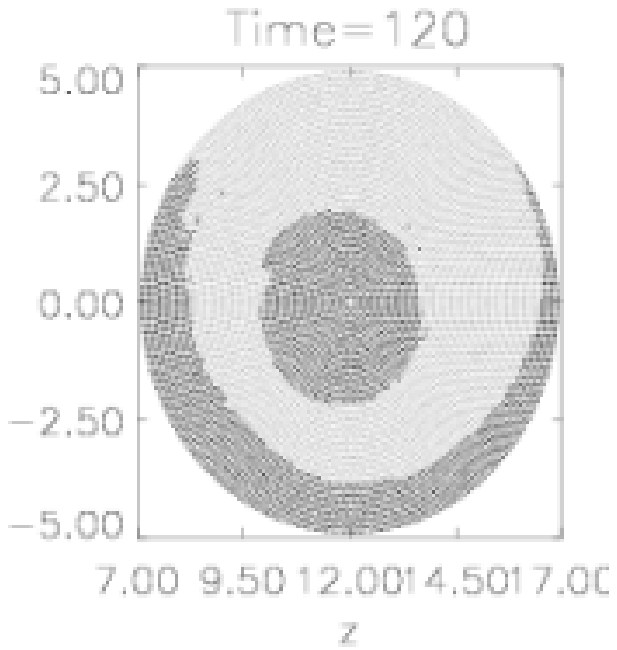}
\includegraphics[scale=.35]{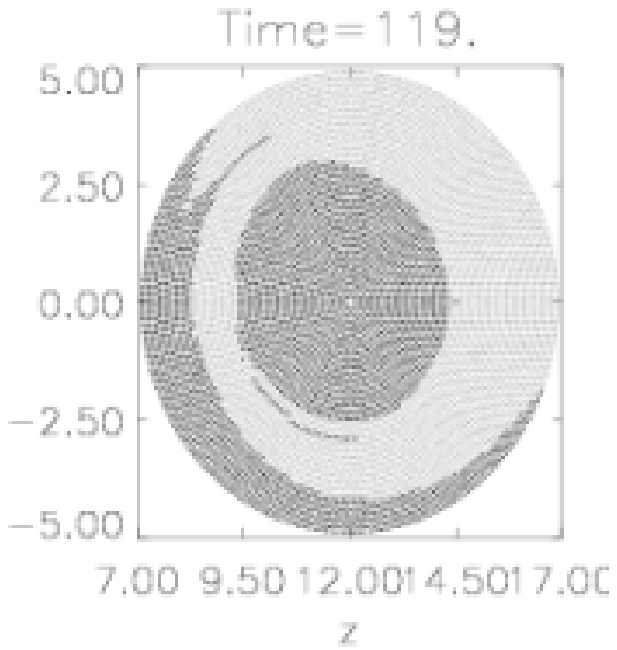}
\includegraphics[scale=.35]{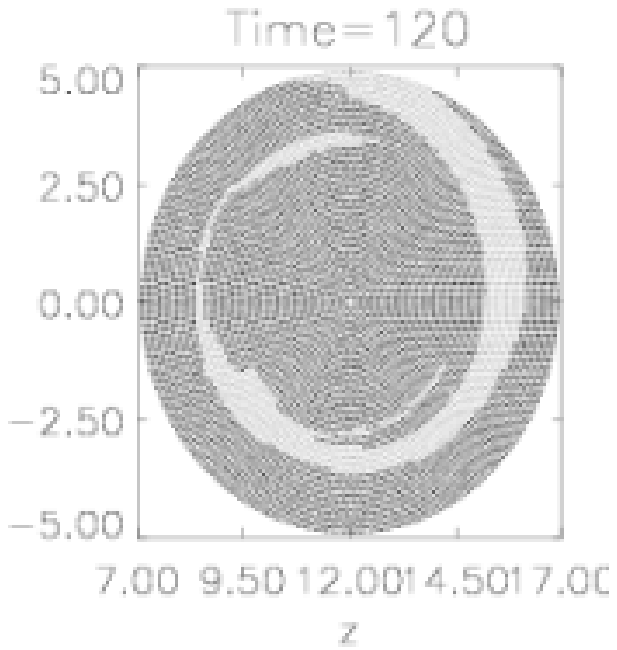}
\includegraphics[scale=.35]{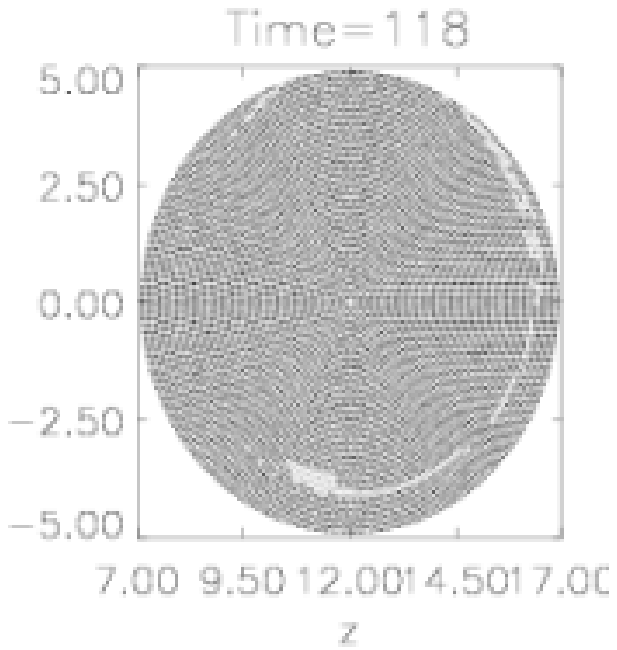}\hfill}

\caption[]{\label{connectivity_plots.fig} Connectivity plots for the five experiments. 
The five columns represent the A, B, C, D and E case respectively.
The rows show the connectivity at times close to  60, 80, 100 and 120.
The disks show the area in the tube at the $y-$boundary from where the fieldlines 
are initially traced.}
\end{figure*}

Figure \ref{connectivity_plots.fig} consists of five columns. Each column 
corresponds to a different experiment and shows how the field lines, 
which have been traced from inside the selected disk, change their connectivity 
in time. The left column, for example, shows the connectivity for experiment A. 
The four panels in the left column show that the connectivity changes first at the 
outer layer of the disk and then moves toward the center following a swirling 
motion. This is because the fieldlines which are traced from the outer periphery of the 
disk, reach higher levels in the atmosphere and reconnect first with the ambient field. 
At the end of the simulation only few fieldlines, which are found in a short distance 
around the center of the disk, have not changed connectivity yet. All the other fieldlines 
have already been reconnected with the ambient field.

We also find that at each time the number of fieldlines that do not change connectivity 
is smaller, as the relative horizontal angle between the two flux systems increases 
from $\phi_0=0$ (right column) to $\phi_0=180$ (left column). As a result, there are very 
few fieldlines that have been reconnected at the end of the simulation for experiment E, 
whereas most of the fieldlines inside the disk have change their connectivity in experiment A.

It has to be noticed that the disk used for tracing the field lines is not changed in time, 
implying that the starting points do not represent exactly the same footpoints in time. 
Investigating the drift velocity of the flux pattern it is found to be far to small to account 
for the change in
connectivity between $t = 60$ and $t = 80$. A closer examination of the field line 
structure shows that field lines at the top of the reconnection sheet continuesly changes
connectivity (as shown in \cite{archontisetal05}), being slowly pushed towards the flanks of
the diffusion sheet. Here they re close
and again become part of the flux tube. In other words, flux from footpoints at side of the
emerging flux region is involved in reconnection processes more than ones already in the
early phase of the emerging process.


\begin{figure}[!htb]
\centering \includegraphics[scale=.48]{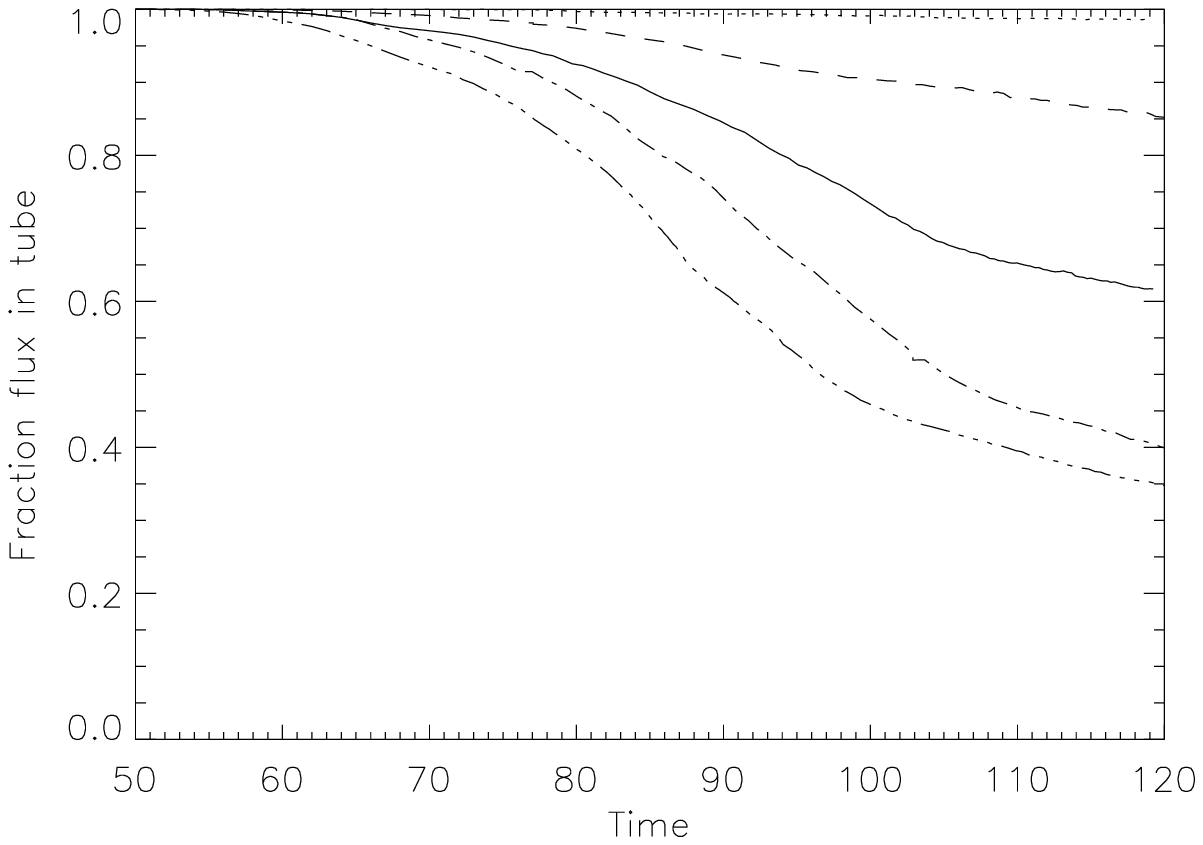}
\centering \includegraphics[scale=.48]{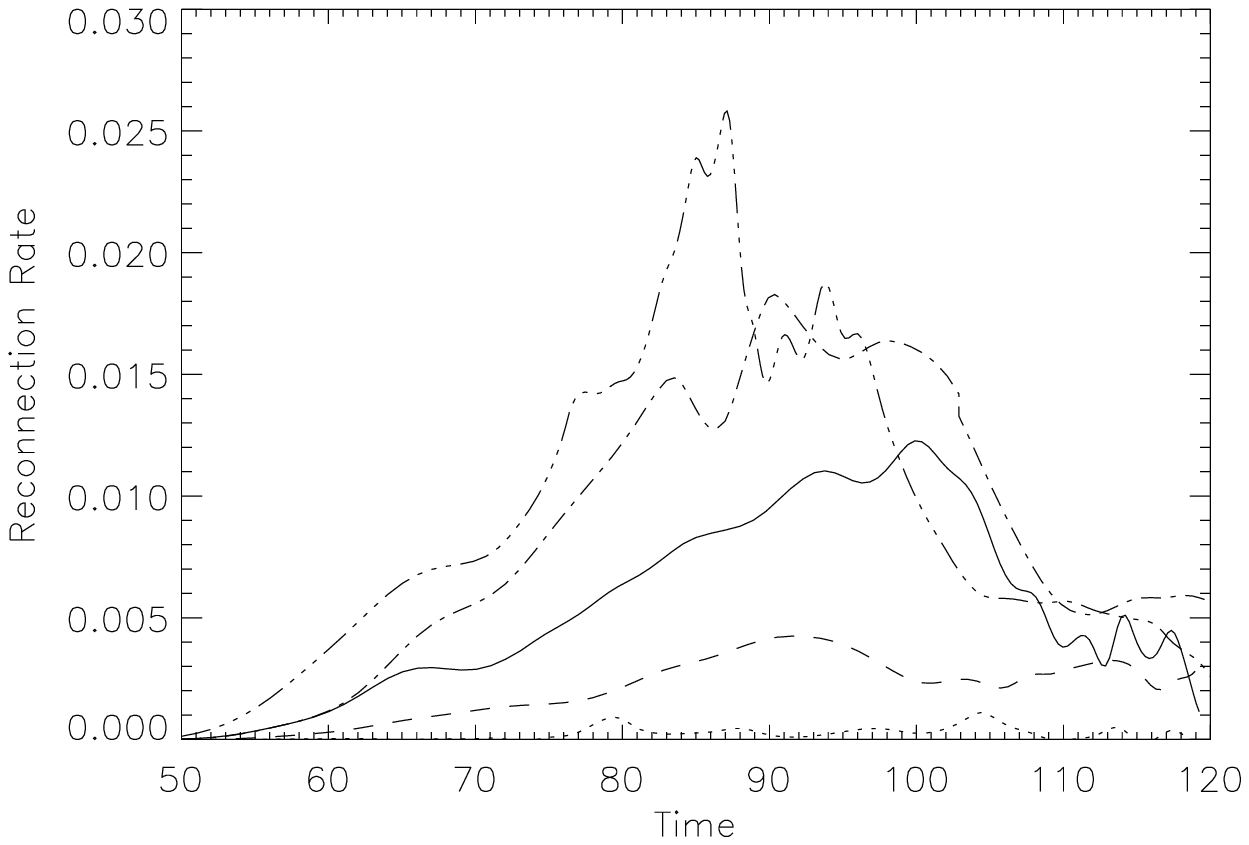}
\caption[]{\label{connectivity.fig}
The left panel shows how the flux connectivity, as defined in \Eq{connect.eq}, 
changes with time for the different experiments.
The right panel represents the reconnection rate of the experiments
by simply estimating the gradient of the curves in the left panel.
The line style is given in Table \ref{names.tab}.
}
\end{figure}
\subsection{Flux connectivity} \label{flux_connectivity.sec}
To further illustrate the results obtained in the previous section, 
we calculate the amount of flux that remains in the tube and does not reconnect, 
normalized to the total flux within the disk, as a function of time. 
\EQ
\label{connect.eq}
\Phi(t) = \frac{\int_{\hbox{black area}} B_{y}(x, l, z, t) dx dz}{\int_{\hbox{disk}} 
B_{y}(x, l, z, t) dx dz}.
\EN
$\Phi(t)$ is shown in the top panel of \Fig{connectivity.fig} for the five experiments 
in \Tab{names.tab}.

\begin{figure}[!htb]
\centering \includegraphics[scale=.4]{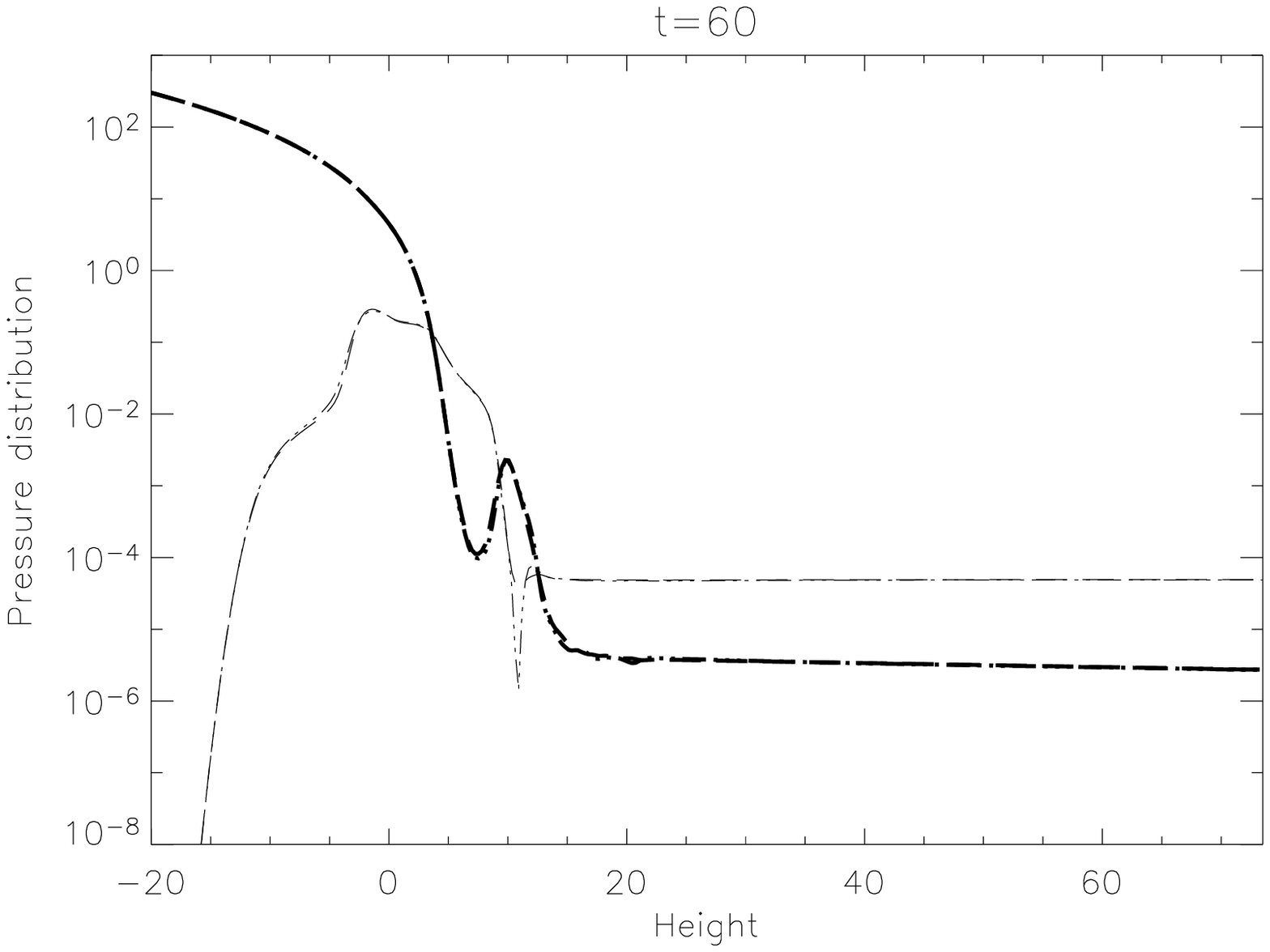}
\centering \includegraphics[scale=.4]{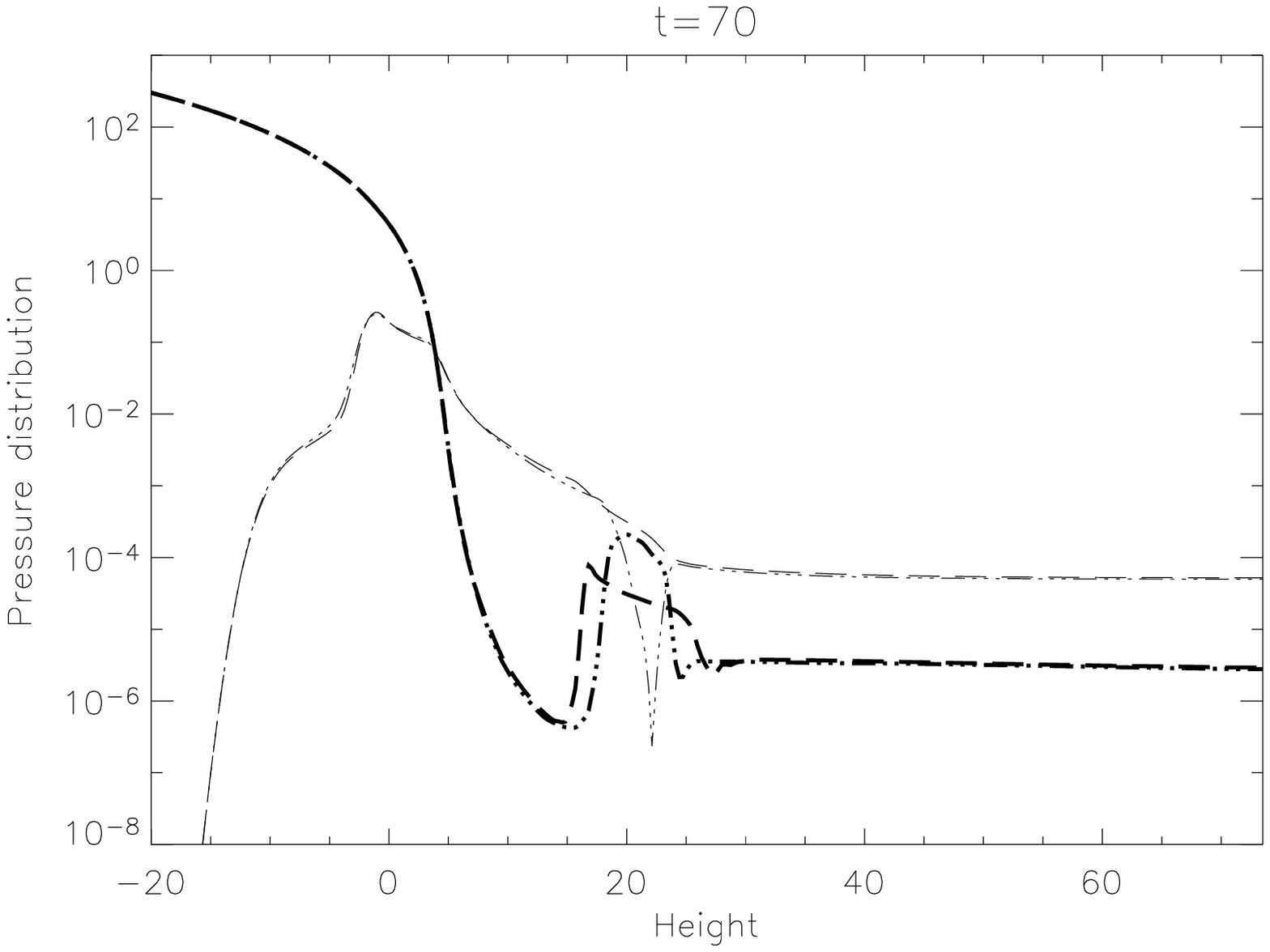}
\centering \includegraphics[scale=.4]{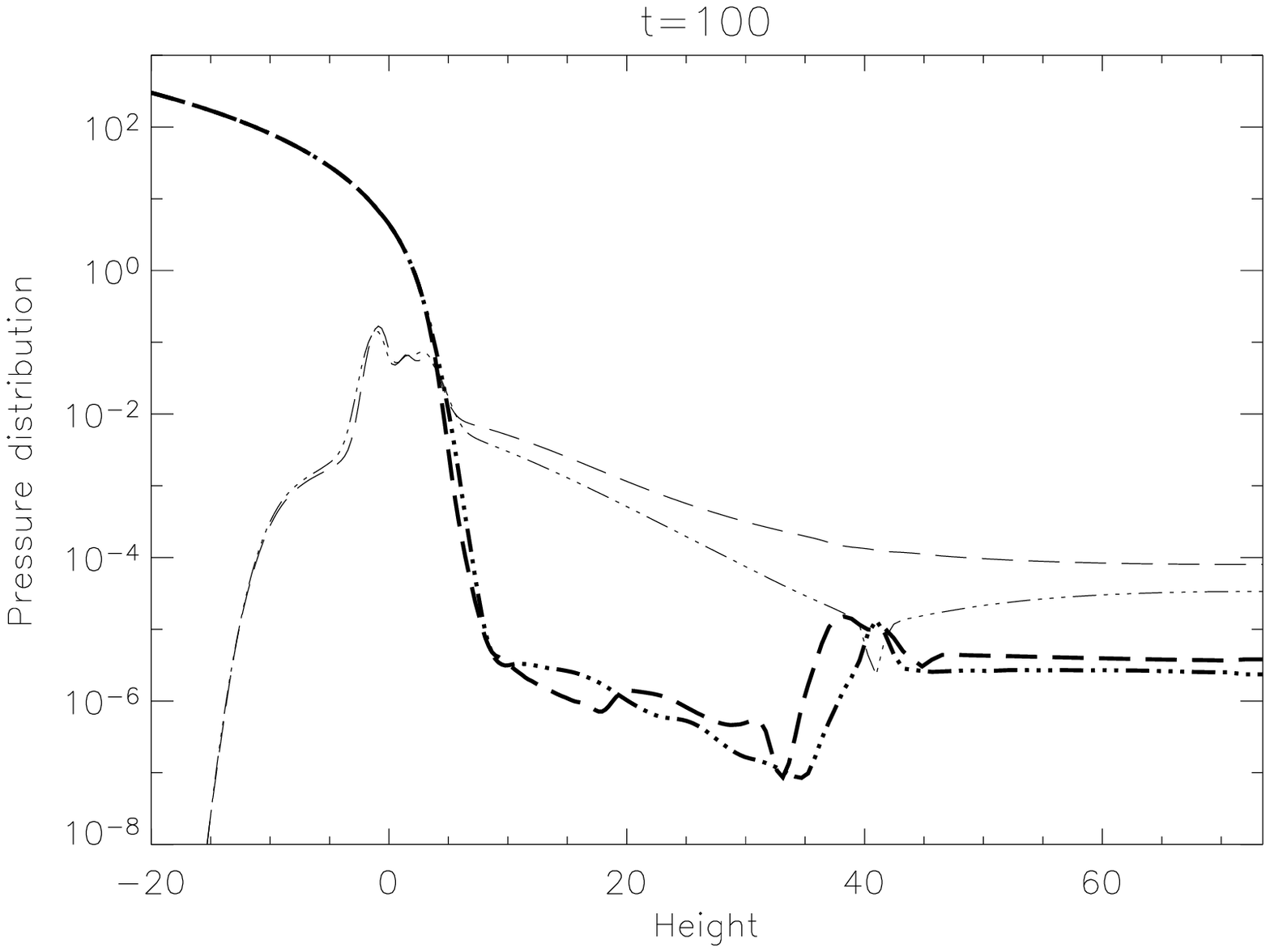}
\caption[]{\label{pressure_height_size.fig}
Magnetic pressure (thin lines) and gas pressure (thick lines) distribution along 
height for the experiments A and D along the central vertical line. Line style are
given in \Tab{names.tab}}
\end{figure}

Flux emergence through the photosphere starts at around $t=60$
and reconnection between the two flux systems starts shortly after this for most of
the experiments. A measure of the reconnection rate is given by the absolute value
of the slope of the lines in the left frame of \Fig{connectivity.fig}. This shows that
there is a short initial phase where the reconnection rate builds up, 
followed by a period of time where the reconnection proceeds
with different, but almost constant, rates in all experiments. After $t \approx 100$ the 
reconnection rate decreases for the three fastest reconnecting experiments, to a lower
level that is roughly maintained until the end of the experiments. 
The reconnection rates are simply given by the gradient of the connected flux fraction.
As already stated earlier, an unknown amount of flux reconnects more than once, implying 
that global-double-separator bifurcations may take place, \cite{Haynes_ea07},
through which recycling of the flux may occur, \cite{Parnell_ea07}.
The estimates of the reconnection rate are, therefore, only providing a minimum value and
cannot be used, to estimate the reconnection speed in the reconnection processes. This 
implies that a quantitative comparison with \cite{Longcope_ea05} results is not possible.
\cite{Longcope_ea05} used TRACE and MDI observations to estimate the energy transfer
between a new emerging region and old coronal magnetic field. Their results showed 
that reconnection was not active for a long initial phase, after which a large fraction 
of the emerging flux connected to the coronal field over a relative short timescale.
In our simulations, we find a more smooth increase in the flux interaction, with a
clear leveling off towards the end of the experiment. Hidden in this may well be a 
significant restructuring of the field that is not apparent due to limitations in 
following the connectivity of individual flux concentrations.

Despite this, the graphs still provide information about the general development 
of the experiments.
The right frame of \Fig{connectivity.fig} indicates that experiment A has the highest 
rate of reconnection and that the reconnection rate decreases as the angle between 
the two flux systems becomes less favorable and the magnitude of the reconnecting field
component decreases. This fact is unlikely to be changed since the multiple reconnection 
depends on field lines already having changed connectivity once.
Thus, the amount of reconnected flux in experiment A increases up to 65\% by the end 
of the simulation while it remains close to zero when the two flux systems are parallel. 
Finally, we find that the amount of flux that remains in the tube at the end of the 
experiments (t=120) decreases nearly as $1 - 0.65 \sin(\phi_0/2)$. In other words, it 
scales with the orientation, and therefore the strength, of the current sheet.

\section{Dynamics of emergence}
\label{dynamics.sec}

In the previous section we showed that the height-time relation of the apex of the tube is 
similar in all experiments. It is also found
that the amount of flux that emerges through a certain height as a function of time 
does not depend on the orientation of the ambient field. However, the amount of flux that 
changes connectivity between the emerging flux and the coronal magnetic field, depends critically 
on the relative orientation of the two flux systems. 

In the experiments where the rate of change of connectivity is high, the overlying 
coronal magnetic field is constantly removed by the reconnection process. Thus, the 
volume above the rising tube is opened and the buoyant system can make its way up 
into the upper atmosphere. On the other hand, in the experiments with very
little reconnection, the coronal magnetic field is not easily removed but instead is 
pushed upwards and keeps up resistance to the rising motion of the tube (the fieldline 
topology at the top part of the emerging tube is illustrated in \Sec{reconnect.sec}).
Thus, one may ask why the rising motion of the 
tube is not influenced by the change of connectivity in the different experiments.

In fact, the height-time relation of the apex of the tube indicates that the process of flux 
emergence is predominantly governed by the dynamics of the rising magnetized plasma.
Thus, in the following sections, first we consider the gas and magnetic pressure
distribution along height, inside the expanding rising volume and across the current sheet,
for two experiments with different initial relative angle (\Sec{pres_dist.sec});
then we study the temporal evolution of forces that act on the upper part of the buoyant 
tube (\Sec{forces.sec}).


\subsection{Pressure distribution}\label{pres_dist.sec}

It has been shown, (Fig.5 in \cite{archontisetal05}), that the three-dimensional 
current sheet which is formed between the two flux systems in experiment A, 
is the location of a rapid change in the direction of 
the magnetic field. In the early phase of the evolution of the system the total magnetic 
field vector goes through a tangential discontinuity across the current sheet with a clear 
minimum at the center of the sheet. As time proceeds the direction of the field changes smoothly 
across the interface of the two flux systems following a rotational-like discontinuity.

Figure \ref{pressure_height_size.fig} shows the gas pressure and magnetic pressure 
along the central vertical line for the experiments A and D at $t=\, 60,\, 70$ and $100$.
The first panel of \Fig{pressure_height_size.fig} ($t=60$) shows that in both
experiments
the magnetic pressure is higher than the gas pressure inside the expanding volume
($5<z<10$) by almost two orders of magnitude. The plasma $\beta$ in this region is
therefore very low.
The magnetic pressure decreases across the interface between the two flux systems 
and has a minimum value inside the current sheet. This is most easily seen in the
top left panel of \Fig{pressure_height_size.fig} for the triple dotted dashed line.
This position corresponds to the pronounced minimum of the magnetic pressure occurring 
at the position of maximum electric current, that is due to the tangential 
discontinuity across the sheet. At the same time the total pressure has a smooth
change over the current sheet. This implies that the plasma $\beta$ increases 
in the current sheet and becomes larger than unity.
At this early stage of the experiment reconnection at the top of the rising 
tube has not started yet and thus the pressure distribution is almost identical in the 
two experiments.

The top right panel of \Fig{pressure_height_size.fig} shows the pressure distribution 
when reconnection occurs between the tube and the ambient field. 
In experiment A, the magnetic pressure still goes through a sharp minimum, although
the value of the minimum is higher than at t=60. In experiment D, instead, then
magnetic pressure has a smooth distribution across the interface, with no minimum.
In either case, the gas pressure supplements the magnetic pressure across the 
interface, so that the total pressure distribution has no extrema here. The reason for
the different behavior of the magnetic pressure profiles for the two experiments is
the presence of an important non-zero, non-reconnecting field component in the 
current sheet in experiment D which is nearly absent in experiment A.

Finally, the bottom panel shows the distribution at a later time, at $t=100$. 
Now, the direction of the total magnetic field vector in experiment A changes smoothly across 
the current sheet following a highly compressed rotational discontinuity. Thus, there 
is a finite non-reconnecting magnetic component in the current sheet and the magnetic 
pressure there becomes higher compared to the magnetic pressure at $t=70$. The gas pressure 
in the top of the rising system has decreased because the dense plasma which was carried 
upwards has been reconnected. In experiment D, the gas pressure and the magnetic pressure 
do not change dramatically compared to the same distribution at earlier times. Thus, 
the magnetic pressure inside the expanding volume in experiment D now is much higher than 
the magnetic pressure in experiment A. At all times, the magnetic pressure exceeds the 
gas pressure and dense material is lifted up against gravity.

By looking at the pressure profiles with height in the bottom panel of 
\Fig{pressure_height_size.fig}, one notice that in experiments A
the current sheet region experiences a pressure minimum. This is a consequence of the
Bernoulli effect associated with the lateral emission of reconnection outflows
from the diffusion region: the gas pressure depletion in the region helps accelerating
the plasma from above and below into the the current sheet, thereby bringing new
magnetic flux in to the reconnection site. 

%
From these plots it is also found that the total pressure increases with time in 
experiment D. This is a pileup effect as the upper boundary is closed and, thus, the growing 
excess pressure cannot propagate through the top boundary.

\subsection{Forces} \label{forces.sec}

In this section, we focus our attention on the forces that act on the upper part of the 
expanding rising volume, just below the current sheet. In the following, we consider experiment A. 
\begin{figure}[!htb]
{\, \hfill
\includegraphics[scale=.4]{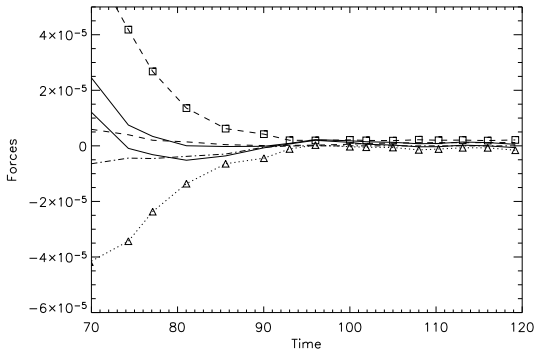}
\hfill \,}
\caption[]{\label{forces1.fig}
Temporal evolution of the vertical forces acting below the current sheet along the ($x=0$, $y=0$)
line. Shown are
the magnetic pressure gradient (dashed-rectangles), the tension force (dotted-triangles),
the gravitational force (dotted-dashed), the Lorentz force (solid), the gas
pressure gradient (dashed) and the total force (thick solid). }
\end{figure}

The vertical component of the 
forces is shown as a function of time in \Fig{forces1.fig}.
The plot shows that the emergence of the buoyant tube is driven by the magnetic pressure 
force that exceeds all the other forces at the early stage of the evolution.
What is also noticeable is that the magnetic pressure force and the magnetic tension are
equal in magnitude and opposite in sign and that they are substantially larger than either 
the pressure gradient or the gravity force.
\begin{figure}[!htb]
{\, \hfill
\includegraphics[scale=.4]{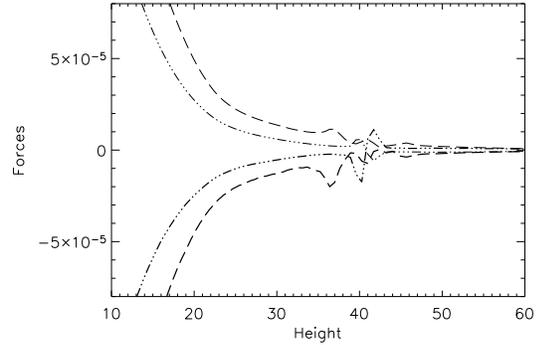}
\hfill \,}
\caption[]{\label{forces3.fig}
The variation of the magnetic pressure force (thin lines) and tension force (thick lines) 
with height for experiment A (triple dotted dashed lines) and D (dashed lines) at $t=100$.}
\end{figure}

The total force is clearly positive accelerating the emerging field against gravity until
$t \approx 72$. This is followed by a short period of deceleration. After $t=95$, all
the forces become very small and are essentially in balance. This result is consistent with the 
motion of the plasma, which rises with an almost constant velocity after $t=95$ (see 
right panel in \Fig{loop_height.fig}).
Notice that the temporal evolution of the total force corresponds well with the 
motion of the apex of the tube shown in \Fig{loop_height.fig}. Also, the time ($t=95$) 
at which the forces take on very small values corresponds to the time at which pressure balance 
is achieved across the current sheet.

%


An analysis of the forces for the experiments B-D show that the temporal evolution 
of the total acceleration is similar to the experiment A. This result explains why 
the apex of the tube reaches almost the same height at the same time in all experiments. 
To further illustrate the point made above we examine the total force for the experiments 
A and D for $t=100$. The $z-$component of the total force is
\EQA
  \Fz  & = &  \Flz + \Fg + \Fp,
      \label{eq:totalforce}
\ENA
where $\Fg=-\rho g$ is the gravitational force, $\Fp=-\dPz $ is the gas pressure force 
and $\Flz$ is the vertical component of the Lorentz force. The latter is written as
\EQA
  \Flz  & = &  {1 \over 4 \pi}\; (\BB \cdot \nabla)B_{z} - {1 \over 8 \pi}\; \dbbz,
      \label{eq:lorentz}
\ENA
where the first term describes the magnetic tension and gives a force when the 
fieldlines are curved, while the second term is the magnetic pressure force that 
acts from regions of high magnetic pressure to regions of low magnetic pressure.

\begin{figure}[!htb]
{\, \hfill
\includegraphics[scale=.4]{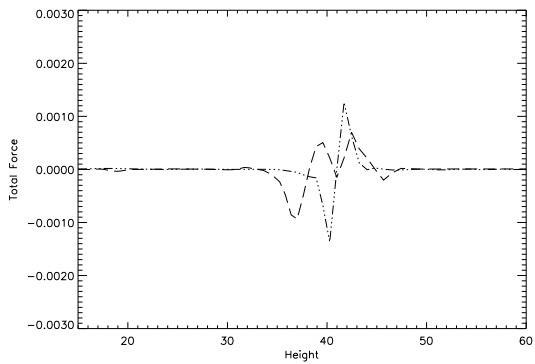}
\hfill \,}
\caption[]{\label{totalforce.fig}
Total force as a function of height for experiment A (tripe dotted dashed line) 
and D (dashed) at $t=100$.}
\end{figure}

\Fig{forces3.fig} shows the distribution of the two terms in \Eq{eq:lorentz} 
along height for the experiments A and D for $t=100$. On the one hand, the vertical component 
of the magnetic pressure force below the current sheet ($z<40$) is larger for experiment D. 
On the other hand, the tension force, which is a downward force, is also larger for experiment 
D and, thus, the Lorentz force has comparable size in both experiments.

Finally, \Fig{totalforce.fig} shows the distribution of the total force in 
\Eq{eq:totalforce} along height for the experiments A and D for $t=100$. The total 
force at the top part of the rising flux system is very small and almost identical for A and D.
Thus, it seems that the total acceleration that acts on the expanding rising volume below the 
current sheet does not depend on the structure of the overlying field and as a result the 
crest of the tube reaches almost the same height at the same time 
for experiments with different orientation of the coronal field.

\section{Topology of the emerging region and magnetic reconnection}\label{reconnect.sec}

The topology of the interacting magnetic field is important for understanding the structure
of the emerging high velocity jets and associated
hotter plasma distribution. In this
section, we show how different is the structure of the magnetic field that appears in the corona when 
the relative orientation of the interacting magnetic fields changes between the experiments.

\begin{figure*}[!htb]
\centering 
\includegraphics[scale=.35]{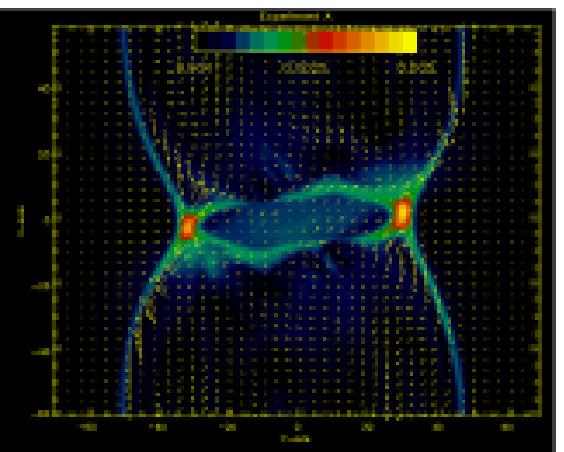}
\includegraphics[scale=.35]{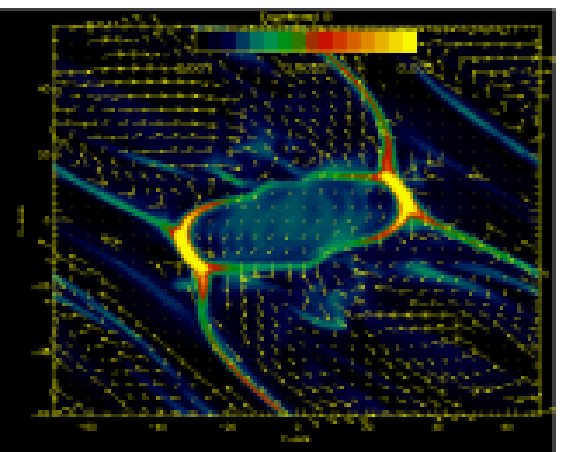}
\centering
\includegraphics[scale=.35]{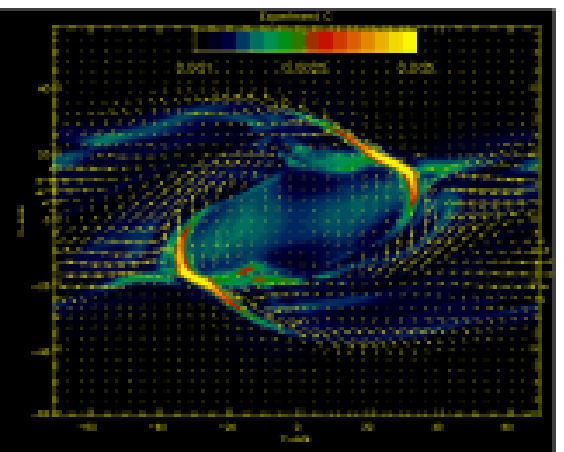}
\includegraphics[scale=.35]{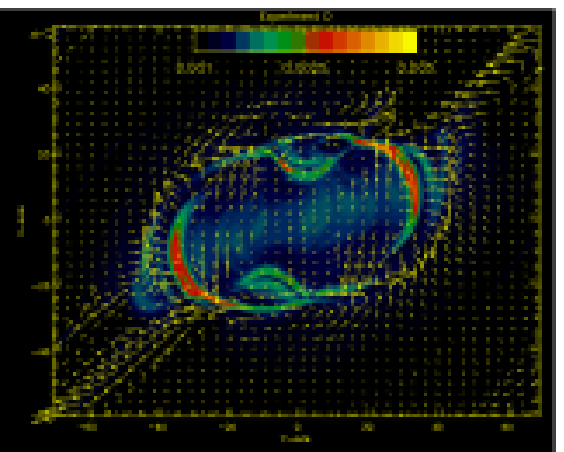}
\centering 
\includegraphics[scale=.35]{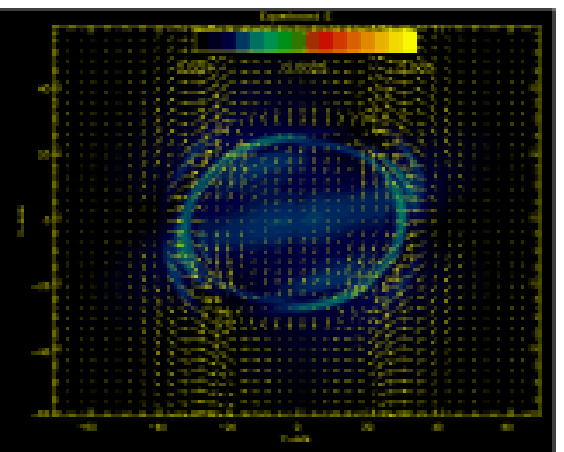}
\caption[]{\label{totalcurrent.fig}
Colourmaps of the total current at $t=100$. Superimposed is the velocity field (arrows). 
The panels correspond to the experiments A-E.}
\end{figure*}

The three-dimensional geometry of the current sheet and the jets, emanating 
from the rims of the current sheet, for experiment A and B, have been studied 
by \cite{archontisetal05} and \cite{galsgaardetal05}. Here, we illustrate the 
projection on horizontal $xy$ planes of the three-dimensional structure of the 
sheet for experiments A-E. \Fig{totalcurrent.fig} shows five panels containing 
colourmaps of the total magnitude of the current ($|J|$) on a horizontal cut at 
a height of 1.7 Mm above the base of the corona at $t=100$. The arrows in the 
panels correspond to the projection of the velocity field. 

The bright patches in \Fig{totalcurrent.fig}, show the location of the highest 
values of $|J|$, and correspond to the 
intersection of the horizontal cut with the arch-like current sheet. Effective reconnection 
occurs and high-velocity outflows are ejected sideways from these sites. The velocities 
of the jets reach values close to 200 Km$sec^{-1}$ in experiment A and B. 
In experiments A-D, the direction of the 
jets is aligned with the direction of the ambient magnetic field. In experiment E there are 
no jets because reconnection does not occur actively between the two flux systems. The arrows 
in this case illustrate the drain of the plasma from the uppermost layers of the emerging 
plasma ball as it rises and pushes the ambient field upward.

The weaker current structure, forming an enclosure between the two locations of
strong current, outlines the border region between the magnetic flux totally connected 
to the emerging flux tube (inside) and either the reconnected flux or the original 
coronal flux (outside). From this it is clear that the horizontal volume of the
emerging flux tube decreases significantly as reconnection becomes more favorably.
It is only the flux contained inside this volume that eventually can end up outlining
the structure of the emerging flux region, while the reconnected field lines connect
the two flux concentrations to neighboring flux regions (here only the corona).

\begin{figure*}[!htb]
\centering
\includegraphics[scale=.27]{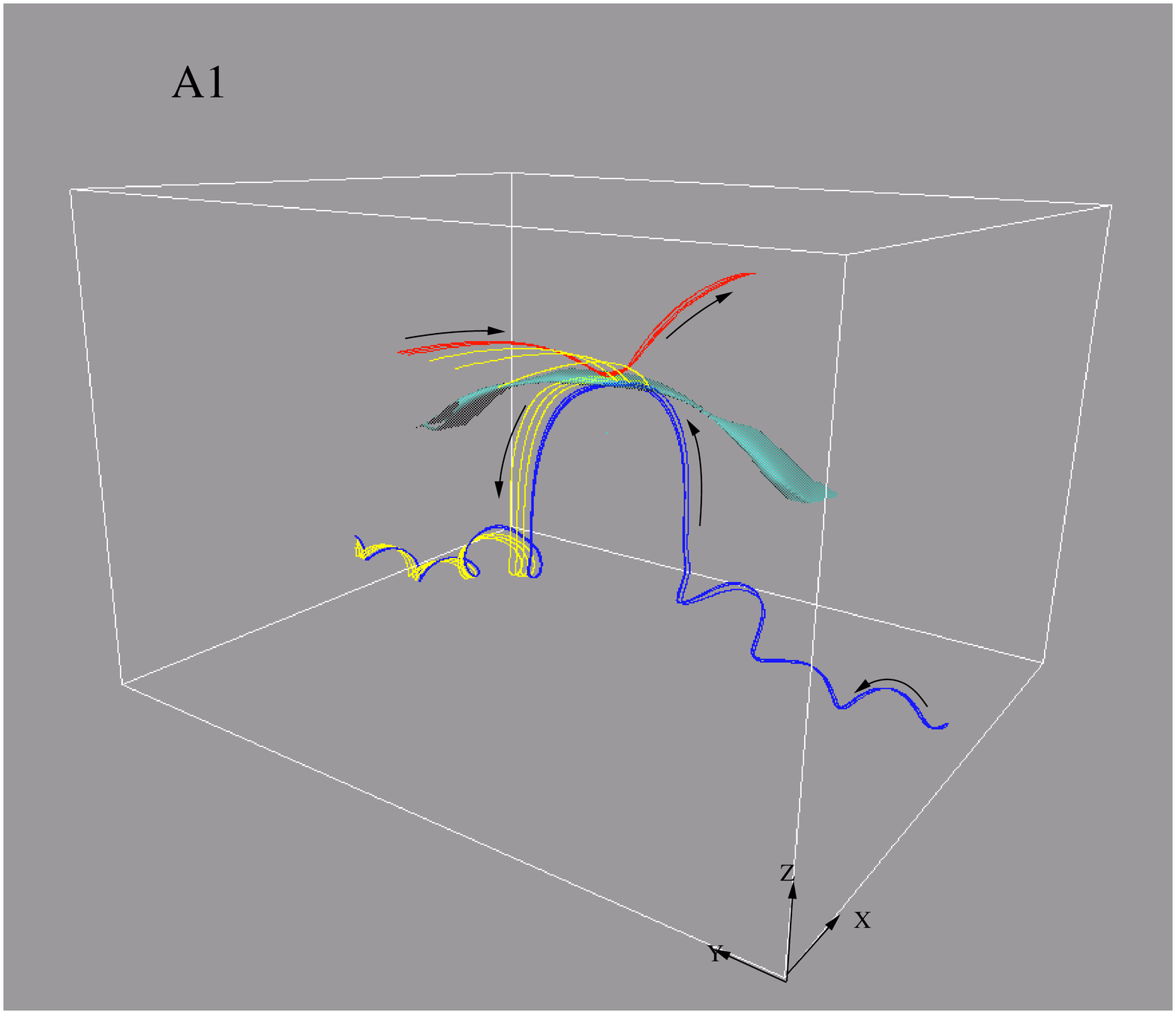}
\includegraphics[scale=.27]{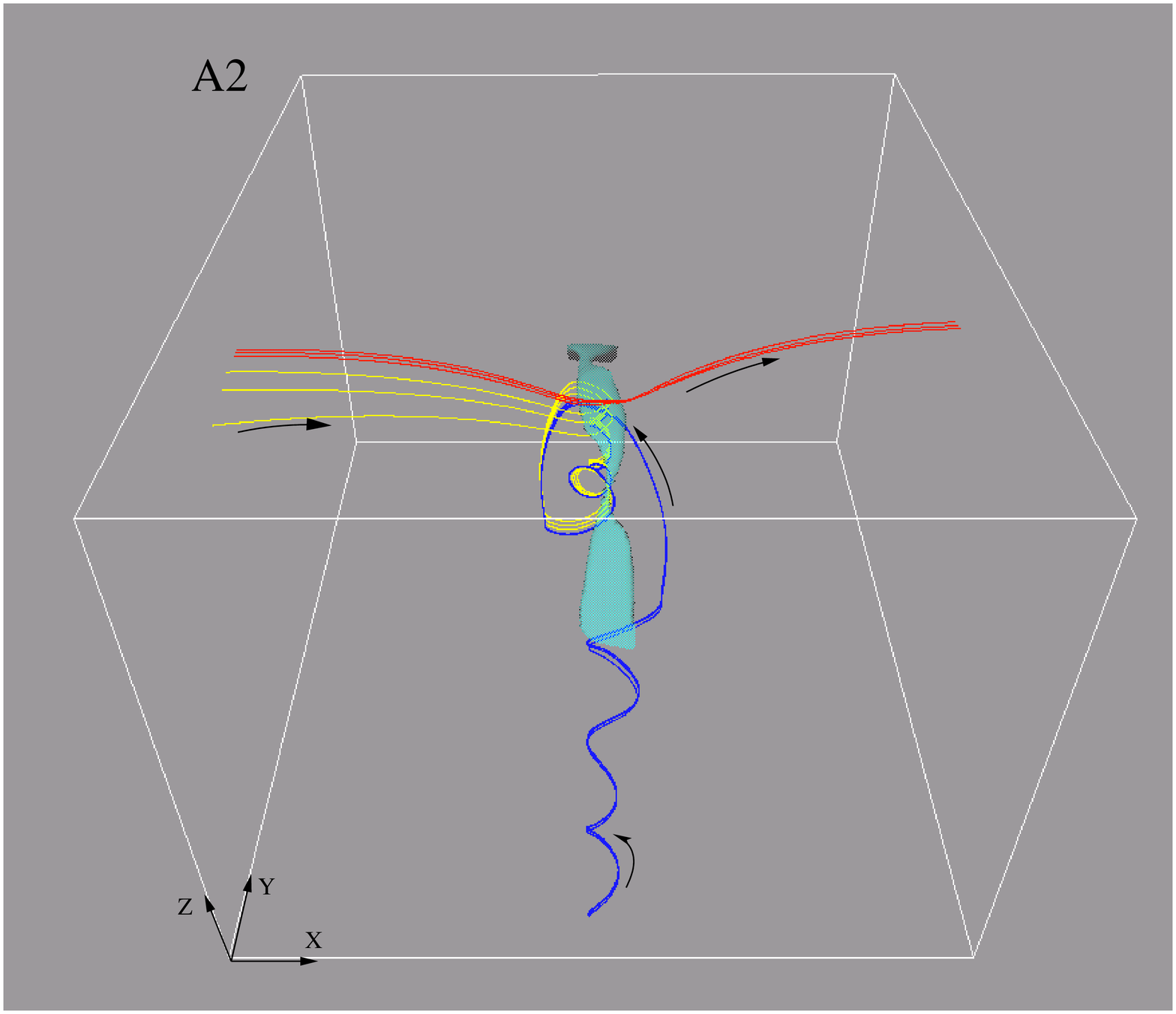}
\centering
\includegraphics[scale=.27]{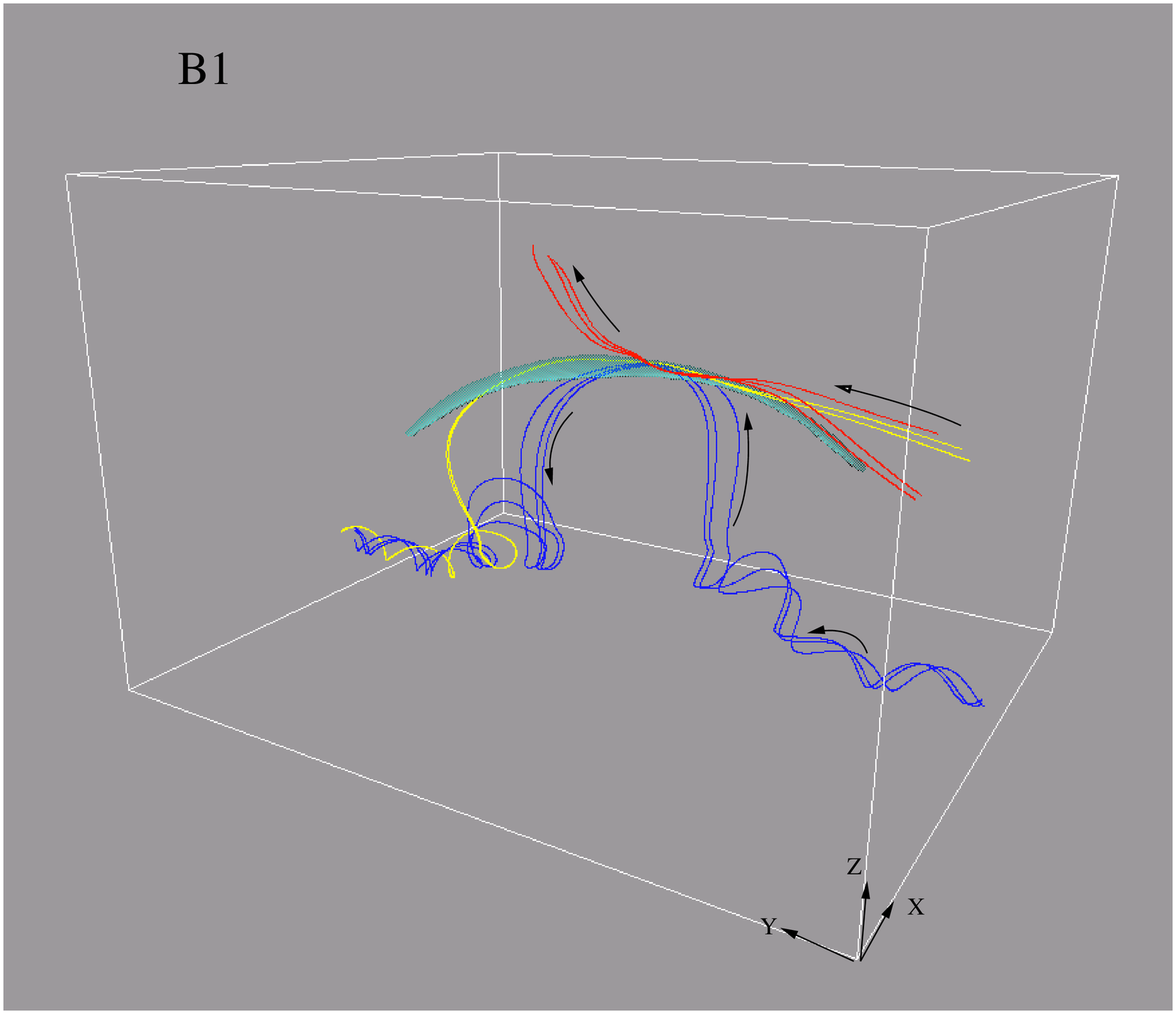}
\includegraphics[scale=.27]{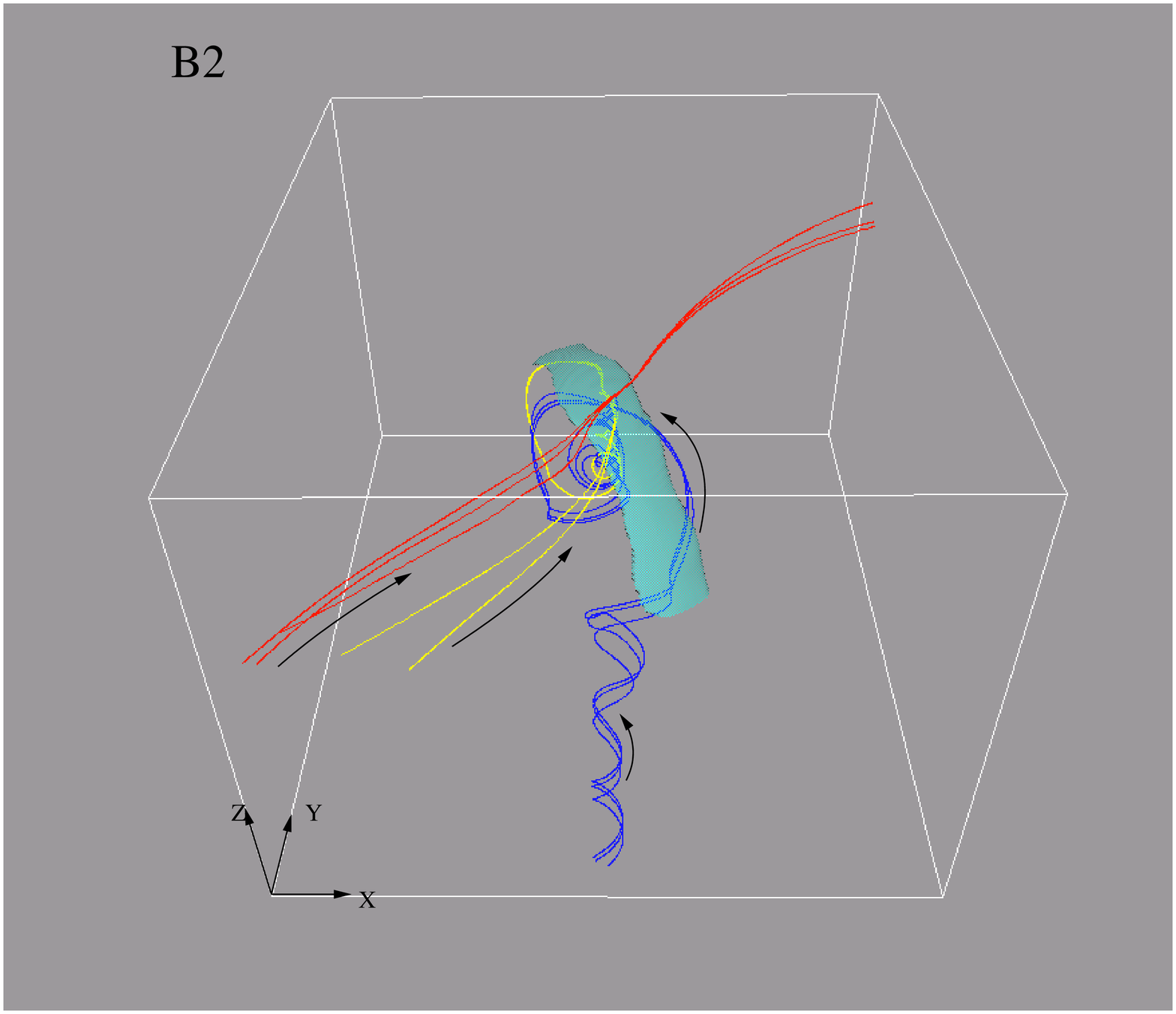}
\centering
\includegraphics[scale=.27]{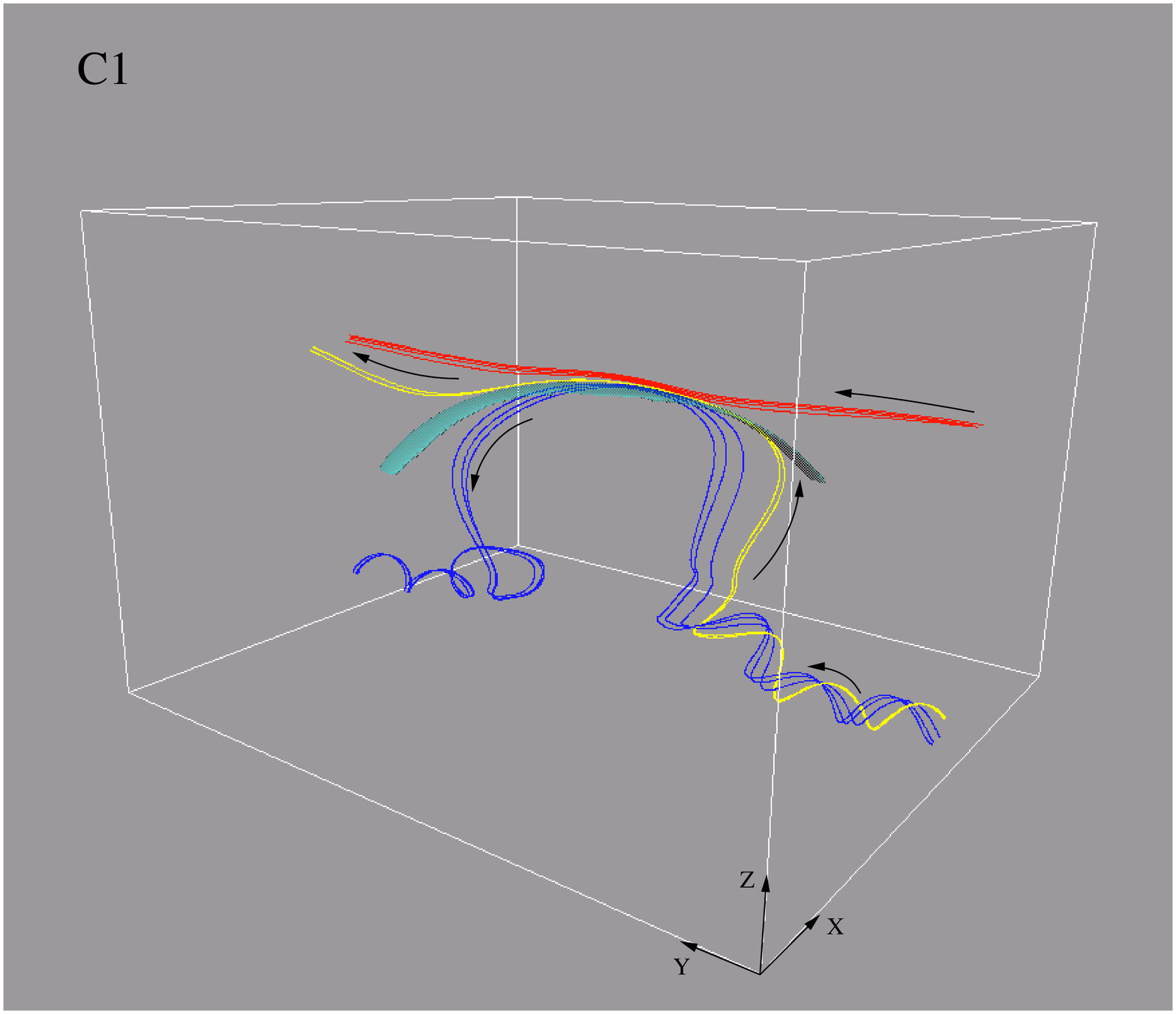}
\includegraphics[scale=.27]{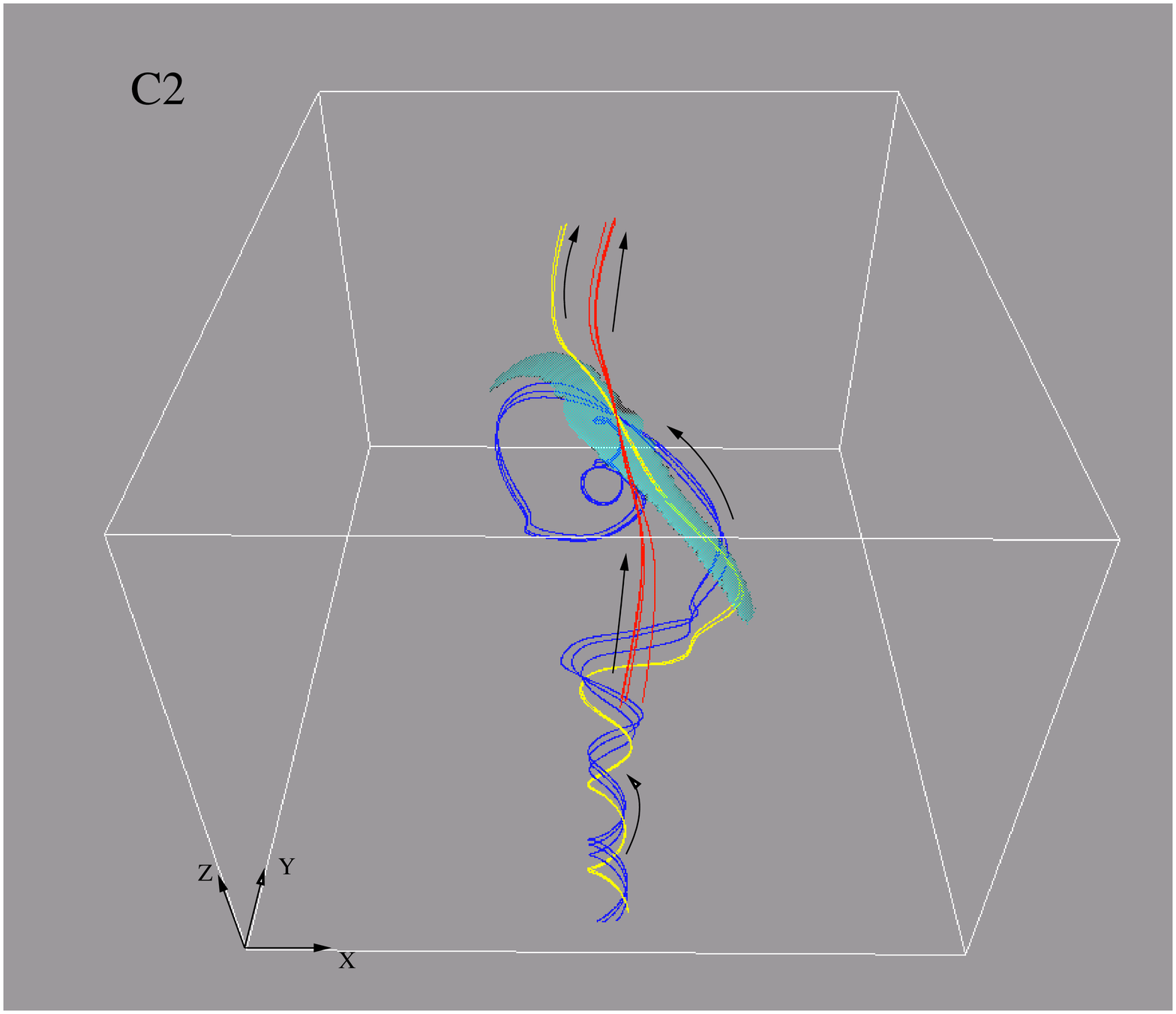}
\caption[]{\label{fieldlines1.fig}
3D visualization of the fieldline topology across the current sheet at $t=100$ for the
experiments A (panels A1, A2), B (panels B1, B2) and C (panels C1, C2). The left column is a 
side view and
the right column is a top view of the same snapshot. The current sheet is visualized as
transparent isosurface. The arrows show the direction of the magnetic field vector.
}
\end{figure*}

\begin{figure*}[!htb]
\centering
\includegraphics[scale=.27]{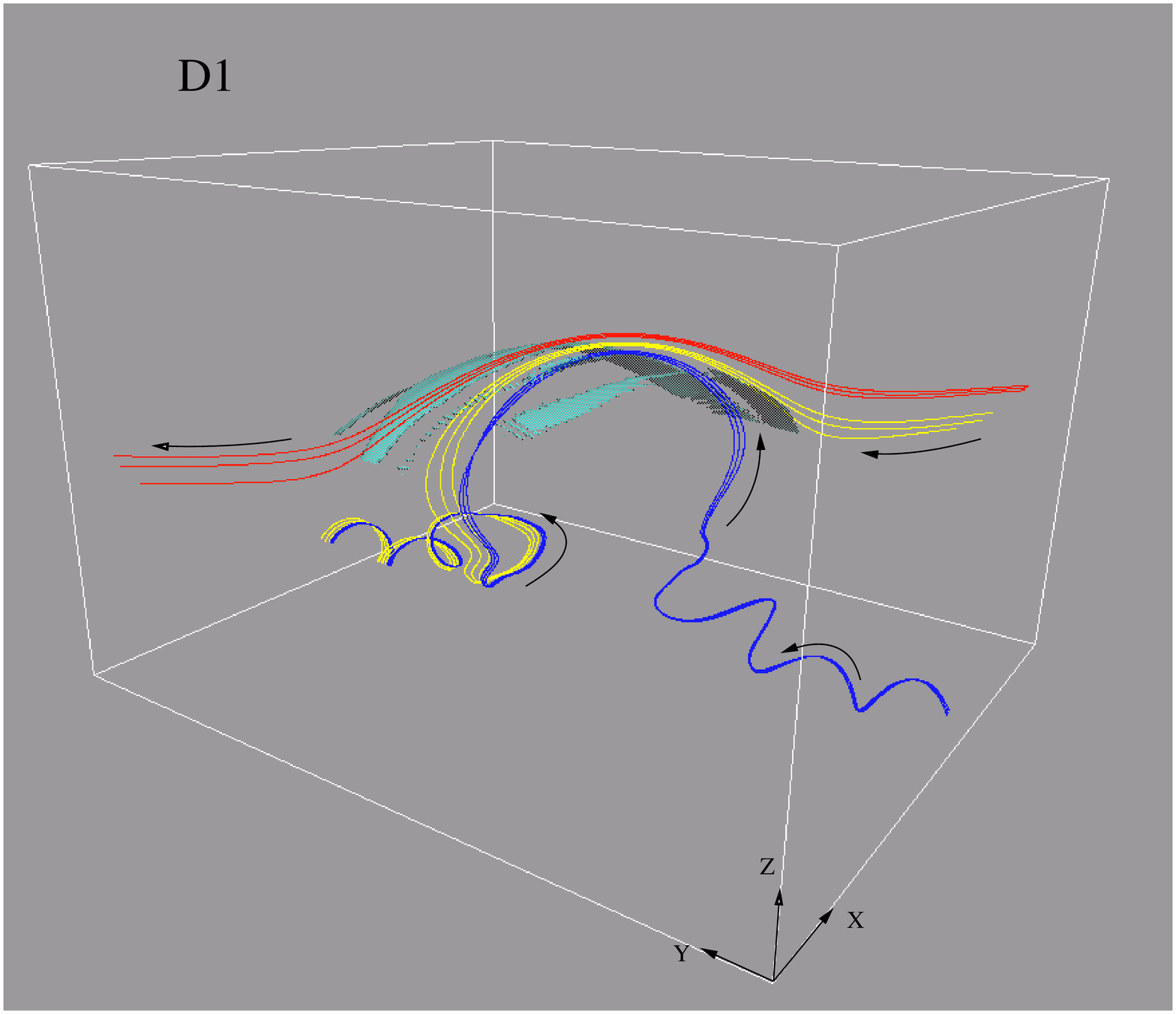}
\includegraphics[scale=.27]{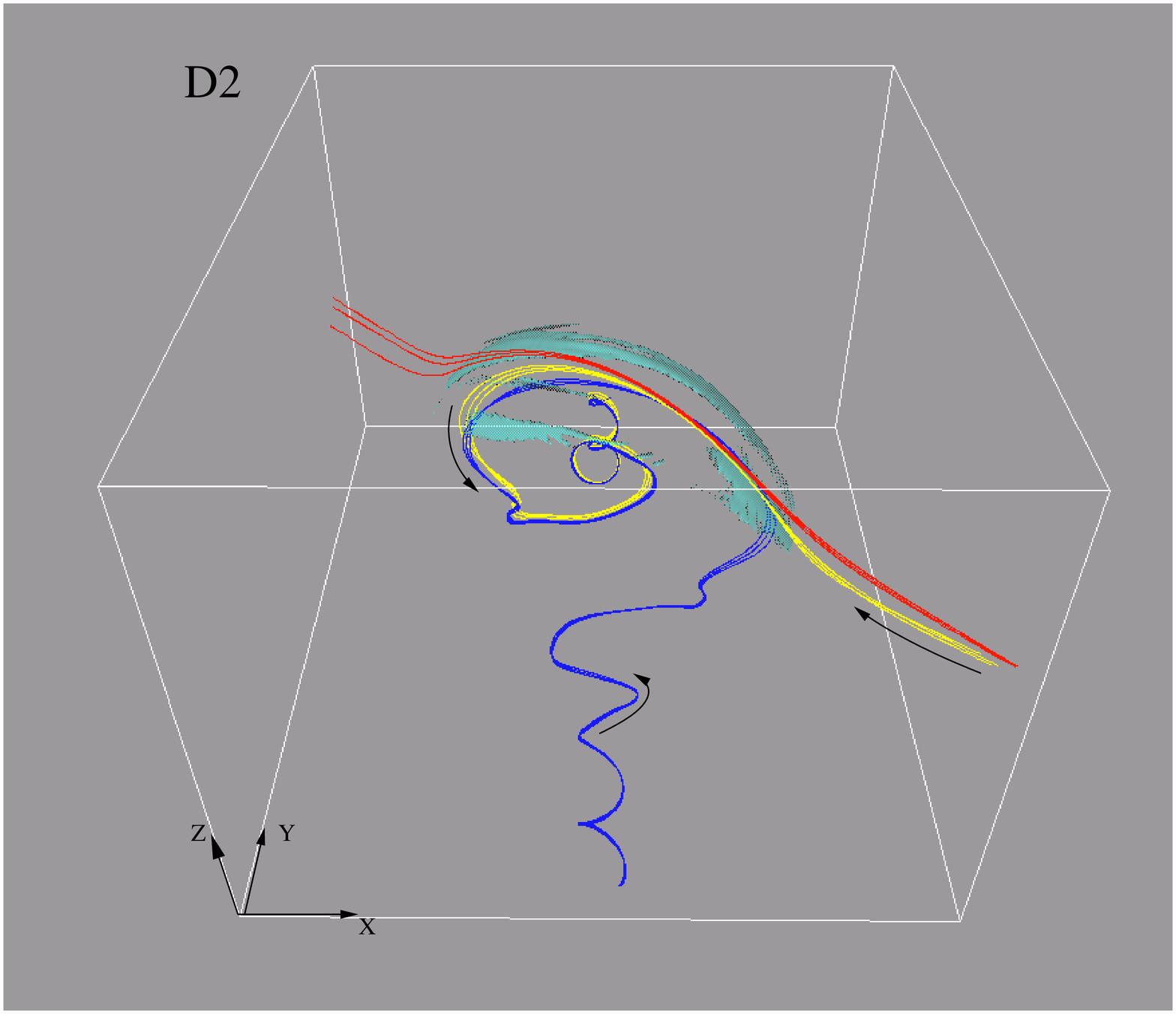}
\centering
\includegraphics[scale=.27]{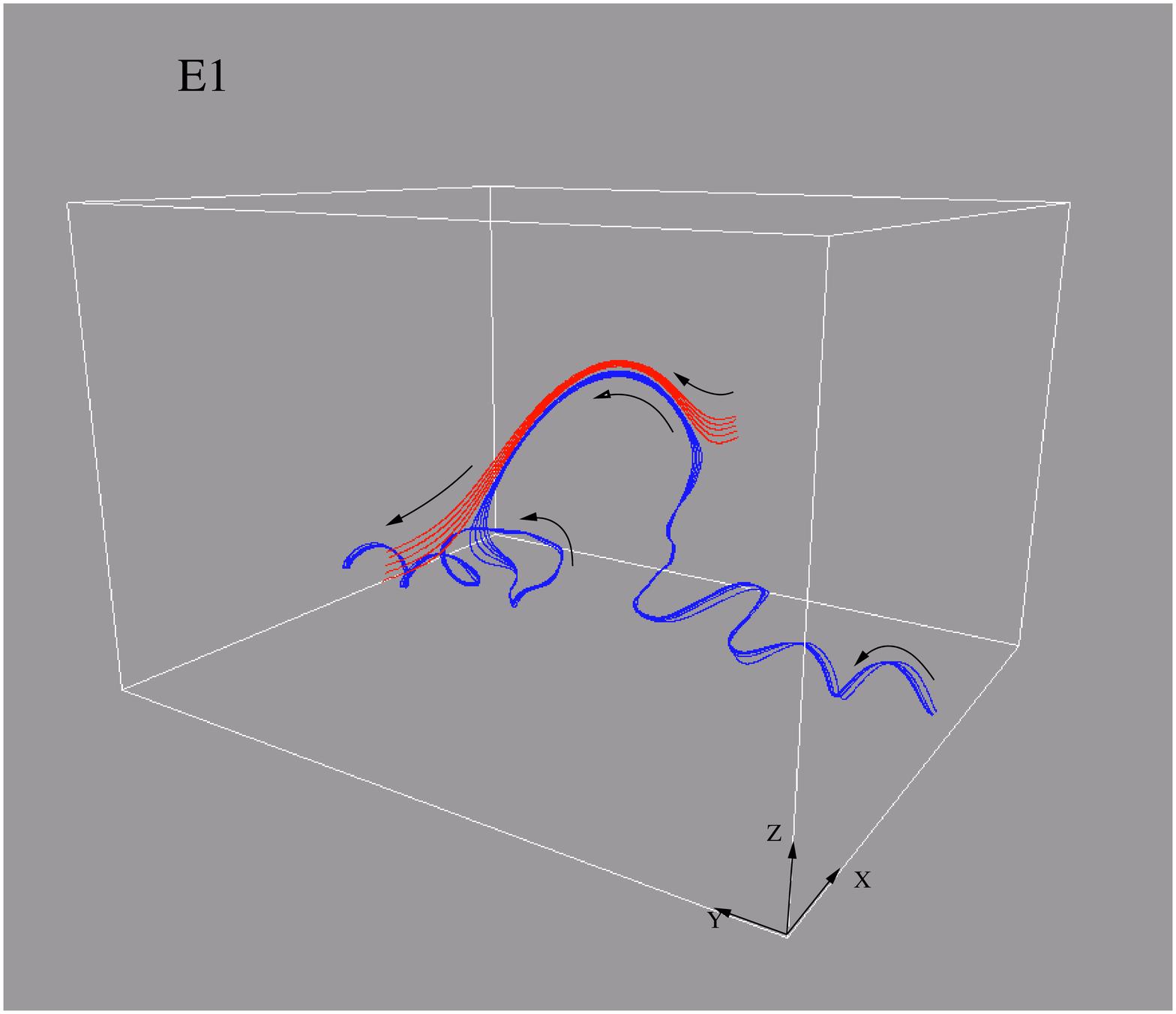}
\includegraphics[scale=.27]{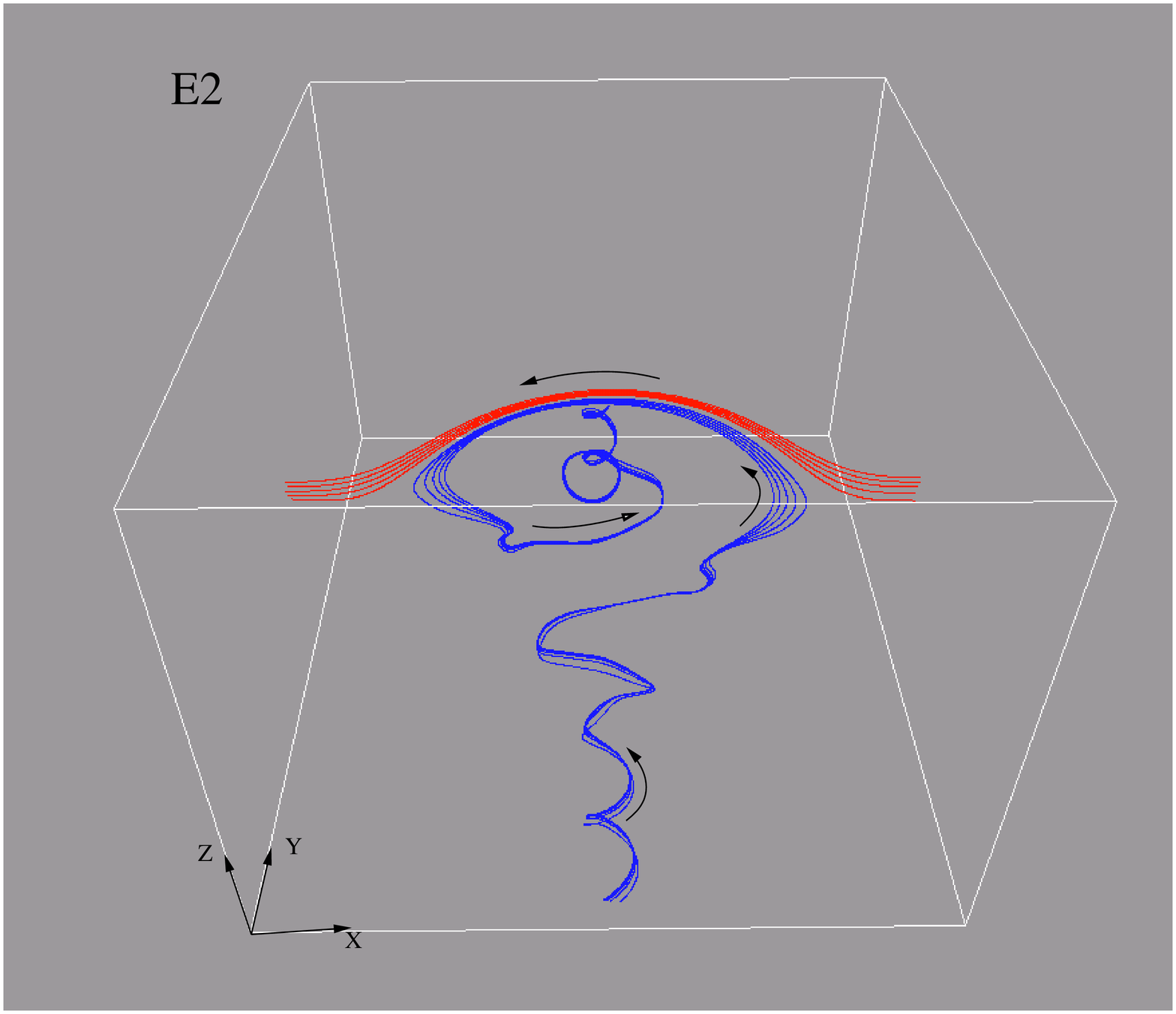}
\caption[]{ Same as in Fig.\ref{fieldlines1.fig} but for the experiments D and E.
\label{fieldlines2.fig}
}
\end{figure*}

To further illustrate how the three-dimensional reconnection works at the top of the 
emerging flux system we study the topology of the fieldlines across the current sheet 
at $t=100$. Figures \ref{fieldlines1.fig} and \ref{fieldlines2.fig} consist of ten panels 
that show the fieldline topology for the five experiments. Three sets of fieldlines have
been traced from different starting positions along height and across the interface of the 
two flux systems close to the center of the emerging region. The blue fieldlines are traced 
from just below the current sheet and belong entirely to the rising tube. 
The red fieldlines are traced from just above the sheet and are ambient fieldlines. Finally, 
the yellow fieldlines are traced from inside the diffusion region and connect the tube 
with the coronal field. The current concentration at the interface is visualized with a 
transparent isosurface. 

The general picture of the reconnection shows a clear difference from the traditional 
two-dimensional configuration. This is because the magnetic field vector across the sheet 
resembles a rotational discontinuity. On the one hand, the uppermost rising fieldlines have 
an orientation which 
is not perfectly aligned with the $x-$axis, as has been explained in \cite{archontisetal05}. 
On the other hand, the orientation of the ambient fieldlines changes from experiment A 
to experiment E so that the relative horizontal angle between the two systems increases.
The product of the reconnection between these two sets of fieldlines is another set of 
fieldlines, the yellow lines, which are ejected sideways from the current sheet and 
establish links between the solar interior and the outer atmosphere. 
Panels B2, C2 and D2 show that the yellow fieldlines at the top of the current sheet have an 
intermediate orientation between the blue and the red fieldlines and that they do not stay
in a two-dimensional plane but they experience full 3D-reconnection. The only experiment where 
the initial relative orientation is not favorable for reconnection is experiment E. Panels 
E1 and E2 show that there is no reconnection (and thus, no yellow fieldlines in the panels).
Instead, the coronal field is pushed upwards and a bended 
hill-like shape interface is formed between the two fields. At the late stages of the evolution 
of the system the ambient field slides down along the sides of the hill of the emerging 
flux and some reconnection occurs at low heights.


Through figures \ref{fieldlines1.fig} and 
\ref{fieldlines2.fig} we also get a confirmation of the dependence of the volume of
the emerged region on the orientation of the coronal field that are shown in 
\Fig{totalcurrent.fig}. More precisely, the blue fieldlines illustrate the geometry of the 
outermost layers of the tube and show that the expansion, in the transverse direction to the axis 
of the tube, is larger when reconnection is less efficient. Indeed, in the case of experiment A 
the outer fieldlines that suffer a large expansion reconnect first and as the time goes on more 
internal layers, which expand less and are also less twisted, come into contact and eventually 
reconnect with the ambient field. In contrast to this, experiment E shows a large expansion of 
the loops of the upper fieldlines as these have not been reconnected due to the small angle 
$\phi_0$. This picture confirms the indications regarding the location of the current structure 
seen in \Fig{totalcurrent.fig}. 

Finally, if we focus on the shape of the emerging fieldlines we find that the blue lines in 
experiment A represent fieldlines which were initially located closer 
to the main axis of the tube and their orientation was not far away from the $y-$axis. When these 
fieldlines emerge (see Panel A1 and A2) keep an almost flat shape in the middle of their crest.
On the other hand, the outermost fieldlines in experiments with less reconnection (see for 
example the panels E1 and E2) represent fieldlines 
which were initially located at the outskirts of the tube and they had larger curvature. 
As these fieldlines rise and expand, they keep their convex shape 
and, thus, the magnetic tension force at the top of the emerging tube becomes larger compared 
to the experiments with more efficient reconnection. 
In fact, this has been also shown in \Fig{forces3.fig}, where the magnetic tension is 
plotted against height for the experiments A and D at $t=100$. 

\section{Discussion}
\label{discussion.sec}
The present experiments were terminated at around t=120 due the use of periodic boundary 
conditions in the horizontal direction. For times after t=120 the effect of these conditions
become very apparent on the experiments as the reconnection jets have propagated across the 
system, and the subsequent evolution is not showing
a freely expanding magnetic flux concentration. In the Sun, this scenario is 
possibly more realistic, than having open boundary conditions. On the other hand, it 
may cause numerical problems and influences the dynamical evolution of the system.
Thus the timescale of each numerical experiment depends also on the boundary effects from the 
periodic conditions. 
 
Some observations seem to favor a situation where a flux rope is emerging into
the corona as coherent structure \citep{Lites_ea95}, while others 
\citep{strouse_zwaan99,Pariat_ea04} indicate a pattern where undulating field lines make
different dynamical effects to release the dense material in the lower
parts along them. In our simulations, including our previous 3D numerical experiments, the initial
flux tube becomes unstable to further expansion into the corona, by building up a 
dominating magnetic pressure force, due to the buoyancy of the tube
\citep{archontisetal04}. When the force becomes strong enough it "blows" the layers close to 
the transition region up and into the corona, where they rapidly push the overlaying material
away and created a dense, cold magnetically dominated plasma dome.
These experiments show that almost 65\% of the
normal flux of the initial flux tube has emerged into the corona. 
We also find that the axis of the initial tube has not fully emerged yet above the photosphere. 
It is wortwhile mentioning that in one of the experiments we find the formation of a horizontal 
current sheet, first reported by \cite{manchesteretal04}, 
which drives internal reconnection of the flux belonging to the emerging tube, allowing 
the lower parts
of the emerging flux tube to disconnect from the emerging flux system. This provides a mechanism 
to decouple the dense photospheric plasma from the field lines that expand into the corona 
(as indicated by \cite{strouse_zwaan99} and \cite{Pariat_ea04}) and provides a possibility 
for forming a structure that looks like an emerged twisted flux tube without emerging
the entire magnetic flux system.

As it is seen from the Sections above, the relative orientation between the emerging
magnetic flux and the coronal magnetic field is of great importance when it comes to the
dynamical evolution of the flux interaction. The emergence process, eventually,
produces high velocity 
plasma jets with temperatures in excess of average coronal values only in the cases of 
efficient reconnection. In less favorable situations the plasma will not be heated much 
and the spectacular display often seen in TRACE movies will not take place. Thus, the 
emergence of an easily observed flux region into the hostile coronal environment depends 
on a relative narrow span of angles between the two systems. On the Sun, this regime may 
be increased compared to the simple model presented here, due to the
much larger structural complexity of the solar environment.

The experiments discussed here relay strongly to the evolution of driven magnetic reconnection.
From analytical investigations of 2D reconnection (see \cite{Priest_Forbes00} and references 
there in), it is found that steady state reconnection depends critically on the structure of 
the magnetic field, the velocity flow and the value of the magnetic resistivity. This makes it 
natural to expect similar dependences to be carry over to steady state 3D reconnection. 
In relation to this, it is obvious that 3D numerical experiments are not able to resolve 
the Reynolds numbers present in the coronal plasma. Why should it then be expected that 
the results from the above mentioned experiments represent an evolution that may take 
place in the solar corona? Using hyperdiffusion the smallest length scale (the 
thickness) of the current sheet is alway only resolved by a few grid points, and it is only 
at these length scales that diffusion of the magnetic field becomes important. Increasing 
the numerical resolution, implies a decrease of the thickness of the current sheet and through 
this a slight delay in the time for the initiation of reconnection. In the driven reconnection
scenario presented here, the largescale of the magnetic field and velocity flow does not change 
significantly when the numerical resolution is increased and, thus, 
the magnetic flux advected into the current sheet will remain approximately the same.
This indicates that magnetic reconnection provides the same new
classes of connectivity. What may change is the local structure of the process, where increased
numerical resolution can allow for reaching local turbulence in the current sheet as it
may become tearing unstable before starting reconnecting. Such a process naturally makes
it possible to reach a more complicated field line connectivity locally, but it also 
implies that the reconnection process becomes independent of the magnetic resistivity and responds
directly to the amount of flux advected into the current sheet 
(\cite{Galsgaard_Nordlund97,Hendrix_ea96,Eyink_Aluie06}). 
Finally, the choise of resolution for the experiments is important.
Experiments with low resolution yield low values for the energy release obtained in the simulations
and large values for the timescale of the evolution of the system. High resolution, especially 
for the volume occupied by the initial tube, may cause problems with the entropy distribution and 
affect the buoyancy of the rising tube.

If the assumption of an emerging magnetic loop is correct, and the processes described in 
this paper and in our previous work are representative
of the emergence of flux from the solar interior to the outer atmosphere of the Sun, 
then there are some simple features that this model may predict.
\begin{itemize}
\item
The process of flux emergence is usually observed using vector magnetograms, which show the 
appearance of bipolar structures at the photosphere. In our experiments, due to the highly 
twisted magnetic flux tube in the subphotospheric layer, the orientation 
of the polarity patterns are, in the early phases, perpendicular to the main axis of the
emerging flux tube (North-South orientation). As time progresses the two opposite polarity 
spots moving apart in the direction along the main tube axis (East-West orientation). 

\item 
The picture of an emerging bipolar region with its strong flux
region located along the axis of the main tube is only true for the photospheric
layer during the short time of these experiments. Slightly above the photosphere
the topology changes and the initial large scale bipolar
region structure going across the magnetic loop is maintained over the remaining time
of the various experiments. Variations between the different experiments are seen, but 
the systematic bipolar patterns going across the main axis of the initial flux tube is
maintained for all experiments even for a height that is only 3400 Km above the photosphere.

\item 
Magnetic reconnection between the emerging and coronal magnetic fields depends on
the relative orientation between the two flux systems. Significant plasma heating and 
high velocity jets will appear when efficient reconnection occurs between 
the two flux systems in a coronal environment. 
For the cases where the relative orientation of the two flux systems is close to parallel, 
a weaker reconnection may take place closer to the photosphere.

\item
The emergence rate seems to depend on the structure of the magnetic field below the
photosphere, which in our experiments has higher field strength than
the coronal magnetic field. The excess magnetic pressure therefore expels the
flux upwards and, thus, opens the volume for a new bipolar structure to appear.
\end{itemize}

The intense reconnection in the near anti-parallel cases makes it possible for most of the
initial flux of the tube to reconnect with the coronal magnetic field. 
If this process continues for long enough, then the
initial connectivity of the flux tube will be totally disrupted. It is therefore possible
that by emerging a bipolar region into the hostile coronal environment, that the initial
connectivity between the two opposite flux concentrations will be, at the least partly,
disrupted. How severe this change in connectivity becomes depends on several factors,
such as the structure of the coronal flux, the amount of the emerging flux
and the duration of the reconnection process. A first estimate of the changes in the 
connectivity may be obtained simply by using potential models.
On the other hand, it seems that these models is not always the best method to follow for the 
study of the time evolution of magnetic structures in actively evolving regions, which is why 
we have not used this analysis here.

Further up in the atmosphere \cite{Longcope_ea05} have analyzed an interesting emergence
observation recorded by TRACE. They adopt the
minimum current model to estimate how much flux has reconnected. Using various assumptions
they find it is possible to account accurately for the changes in connectivity. They also
find that the reconnection between the emerging region and the existing coronal 
flux happens only after some time.  As the interaction starts, it reaches a high activity
level that last for a finite time before rapidly leveling off again.
As most of the involved flux seems to be accounted for, then it
indicates that the system has to build up a significant amount of stress before they start 
interacting. A different possibility is that the lower lying magnetic field structures are 
not favorable for the reconnection process and that the onset of the reconnection process 
does not start until some of the aligned flux has been pushed away and the relative
orientation between the two flux systems has changed. It is also a
possibility that the initial weaker reconnection process takes place in regions where the
plasma density is still so high that the released dissipation can not heat the plasma to
coronal temperatures. Finally there is the possibility that the initial coronal field in
the emergence region is so weak that it does not provide for any significant heating as 
the new flux pushes its way up.

\section{Summary}
\label{summary.sec}

The emergence of new magnetic flux into an existing coronal magnetic field can
evolve in very different ways. The coronal reaction on the emergence depends strongly on the
relative orientation of the two flux systems. Efficient reconnection takes place only in the 
case where the two flux systems have almost antiparallel orientations. Taking 
place in the corona, this provides a 
number of very clear signatures, such as, high velocity jets and intense heating.
The active reconnection process clearly affects the volume occupied by
the magnetic flux totally connected with the underlying magnetic flux tube. 

The reconnection between the two flux systems can slow down mainly because of two 
reasons; Either because all flux in the emerging flux tube has reconnected
with the coronal field, and by this totally disrupted the structure of the emerging
flux rope. Or because the emerging process stops and the two flux systems find a 
mutual balance with each other where no, immediate, reconnection is taking place.
Thus, the length and final structure of the new coronal magnetic field 
strongly depends on the total amount of emerging flux, its ability to reconnect and
on the amount of the coronal magnetic flux.  

For situations where the two flux systems are nearly aligned, it will be possible
for the emerging field to penetrate far into the coronal magnetic field before it
will create any significant coronal signatures revealing its present. These situations
will only be easily spotted if simultaneous observations are made in photospheric
and transition region lines.

In our simulations, we have used a
simple energy equation that does not take into account optical thin radiation or 
anisotropic heat conduction in the corona. These effects are important ingredient
to include in the experiments before more detailed comparisons with observations can be obtained. 
These are effects that will be discussed in future publication. 

In a subsequent paper we are going to discuss in detail the local implications
of the ongoing magnetic reconnection process and its implication on the dynamical
evolution of the different experiments. 

\begin{acknowledgements}
We are grateful for computational time on the UKMHD linux cluster in
St. Andrews (Scotland, UK), funded by SRIF and PPARC, and on the linux
cluster at the IAC (Tenerife, Spain) partially funded by the Ministry of
Science and Technology. This work has also benefited from financial support
through the {\it Platon} European Research Training Network
HPRN-CT-2000-00153 of the European Commission and the CICYT project
no.~AYA2001-1649 of the Spanish Ministry of Science and Technology.
K.G. was supported by the Carlsberg Foundation in the form of a fellowship.
FMI thankfully acknowledges the computer resources, technical expertise and
assistance provided by the Barcelona Supercomputing Center - Centro Nacional de
Supercomputaci{\'o}n¿½½ (Spain)
We thank the anonymous referee for helpful comments that improved this paper.
\end{acknowledgements}

\def\ASPR{{Adv. Spa. Res.}}
\def\AAP{{A\&A}}
\def\SOLP{{Sol. Phys.}}
\def\APJ{{ApJ}} \def\apj{{\APJ}}
\def\APJSS{{ApJS}} \def\apjss{{\APJSS}}
\def\CPC{{Comp. Phys. Commun.}}
\def\GAFD{{Geophys \& Astrophys. Fl. Dyn.}}
\def\JFM{{J. Fl. Mech.}}
\def\PHFL{{Phys. Fl.}}
\def\PHPL{{Phys. Pl.}}
\def\physreve{{Phys. Rev. E}}
\def\PhysRevE{{Phys. Rev. E}}
\def\phpl{{Phys. Pl.}}
\def\GRL{{Geophys. Res. Lett.}}
\def\JGR{{J. Geophys. Res.}}
\def\MEMSAIT{{Mem Soc Astr It}}
\def\PASJ{{Publ. Astron. Soc. Japan }}
\def\ARFM{{Ann.~Rev.~Fluid Mech}}
\def\etal{{\it et al\/}}

\end{document}